\documentclass[aps, showpacs, prd, superscriptadaddress, twocolumn] {revtex4-1}
\usepackage{hyperref}
 \usepackage{amsmath}
\usepackage{amssymb}
\usepackage{epsfig}
\usepackage{natbib}
\usepackage{epstopdf}
\usepackage{graphicx}
\usepackage{floatrow}
\usepackage[caption=false,font=Large]{subfig}
\newcommand*{\be}{\begin{equation}}
\newcommand*{\ee}{\end{equation}}
\newcommand*{\bea}{\begin{eqnarray}}
\newcommand*{\eea}{\end{eqnarray}}

 \DeclareFontFamily{OT1}{pzc}{}
 \DeclareFontShape{OT1}{pzc}{m}{it}%
 {<->  s  *  [1.400]  pzcmi7t}{}
\DeclareMathAlphabet{\mathscr}{OT1}{pzc}%
{m}{it}

\begin{document}

\title{{Understanding}  oscillons: standing waves in a ball}

\author{N. V.  Alexeeva,  I. V. Barashenkov}
 \affiliation{
 Centre for Theoretical  and Mathematical Physics,  University of Cape Town,  South Africa}
 \author{A. A. Bogolubskaya,  E. V. Zemlyanaya  }
 \affiliation{ 
Joint Institute for Nuclear Research, Dubna, Russia
 }

\begin{abstract}
Oscillons are   localised long-lived 
pulsating states in the three-dimensional $\phi^4$ theory.
We gain insight into the spatio-temporal structure and bifurcation of the oscillons
 by studying   time-periodic solutions  in a ball of  a  finite radius. 
 A sequence of weakly localised  
   {\it Bessel waves} --- 
 nonlinear  standing waves
with the Bessel-like $r$-dependence --- 
  is shown to extend from eigenfunctions of 
  the linearised operator. 
  The  lowest-frequency  Bessel wave  serves as 
 a starting point of a branch 
 of  periodic solutions with  exponentially localised cores and small-amplitude tails
 decaying slowly towards the surface  of the ball. 
A numerical continuation of  this branch  gives rise to the
 energy-frequency diagram featuring a series of resonant spikes.
 We show that the standing waves associated with the resonances 
  are born in the period-multiplication bifurcations of the Bessel waves with higher frequencies.
  The energy-frequency diagram for a sufficiently large ball displays sizeable intervals of stability against spherically-symmetric perturbations.

\end{abstract}

\maketitle

\section{Introduction}

Repeated expansions and contractions of spherically-symmetric vacuum domains were 
observed  \cite{Voronov1,BM1}
in computer simulations of 
 the $\phi^4$ equation,
 \be
\Phi_{tt} -\Delta \Phi - \Phi + \Phi^3=0.
\label{A1}
\ee
 More accurate numerical studies \cite{BM2}
  revealed the formation of long-lived pulsating
 structures of large amplitude and nearly unchanging width.

 These structures ---  dubbed oscillons in Ref \cite{G1}   --- 
 have turned out to be of interest in several cosmological contexts, 
 including 
 the dynamics of inflationary reheating,  symmetry-breaking phase transitions, and  false vacuum decay
  \cite{CGM,Riotto,GInt,11cosmo,Dymnikova,Broadhead,bubbling,Amin1,Stamatopoulos,Zhou,Amin2,Adshead,GG,Bond,Antusch,Hong,Cyn,LozAm}.
  Oscillons 
 have  been discovered in the planar Abelian Higgs theory \cite{GT2,Achi}, Einstein-Klein-Gordon equations \cite{Maslov,Zhang2,Nazari,Kou1,Hira,Kou2},
 axion models \cite{Kolb,Vaquero,Kawa_axion,Olle,Miyazaki},  string phenomenology \cite{string,Kasu,Sang}  and 
   bosonic sector of the standard model 
 \cite{Farhi2,Graham,Gleiser4,Sfakianakis}. 
 The oscillon's quantum radiation was evaluated in \cite{Hertz,Saffin}
 and
 the impact of fermionic  corrections  was considered in \cite{Borsanyi}.
 Oscillatory localised structures
 (known as $\mathcal{I}$-balls in that context)
 feature prominently in studies of the adiabatic invariant in theories
 without  electric or topological charge \cite{Kasuya,Kawasaki,Mukaida,Ibe}.
 
  Considerable progress in the understanding 
 of the oscillon properties 
 was achieved  through  the state-of-the-art computer simulations \cite{G1,CGM,Honda,Gleiser10} and numerical
 Fourier analysis \cite{Honda,Fodor1}.
 % Refs \cite{G1,CGM,Honda} focused their attention on the evolution of the gaussian initial conditions with varied radii.
 % By tuning the radius of the gaussian,  oscillons with extremely long (seemingly infinite)  lifetimes were reached \cite{Honda}.
 % These objects  were interpreted as unstable solutions with a single instability mode.
  Most importantly, the authors of Ref \cite{Fodor1} demonstrated the existence of  periodic solutions
  with frequencies filling the entire $(0, \omega_0)$ interval.
  (Here $\omega_0$ is the frequency of spatially uniform small-amplitude oscillations about the vacuum.)
  The solutions in question have
   exponentially localised cores and 
  oscillatory tails,  with the tail amplitudes
   decaying in proportion to $r^{-1}$.
   The authors of Ref \cite{Fodor1} 
 have interpreted the evolution of oscillons as an adiabatic motion in the parameter space of those ``quasibreathers".
 
At the same time,   theoretical arguments
produced estimates for the oscillon's
 energy, radius, frequency, core amplitude, and lifetime \cite{GS1,GS2}.
 These were based on a heuristic combination of linear radiation analysis 
 and a single-mode variational model \cite{CGM,G2,GS1, GS2}.
A refined perturbation expansion of the small-amplitude oscillons \cite{Fodor2} is also worth to be mentioned.
  
  %The aim of the present study is to shed further light on the structure and  nonlinear dynamics of the oscillon.
  %  by considering  periodic standing waves with a node   on the sphere  of a fixed finite radius.   

    The aim of the present study is to shed further light on the structure and resonant properties  of the oscillon
  by examining periodic standing waves in a ball of a large but finite radius.

   To make it more precise, 
  let  $\Phi(r,t)$ be a 
   spherically-symmetric solution of equation \eqref{A1}
   approaching $\Phi_0= -1$ (one of  two vacuum solutions) as $r \to \infty$.
   The difference
\[
\phi=  \Phi-\Phi_0
\]
obeys
\begin{subequations}  \label{2D}
\be
\phi_{tt}- \phi_{rr} - \frac{2}{r} \phi_r + 2 \phi- 3 \phi^2 + \phi^3=0.
\label{A200}
\ee
Instead of searching for  solutions of  the equation  \eqref{A200}  vanishing at infinity, 
 we consider solutions 
satisfying the boundary conditions
%\begin{subequations} 
\be
\label{BC}
\phi_r(0,t)= \phi(R,t)=0
\ee
with a large  $R$. 
{(The first condition in \eqref{BC} ensures the  regularity of the Laplacian  at the origin.)}
One more boundary condition stems from the  requirement of periodicity with some $T$:
\be
\phi(r,    T)=\phi(r,0).
 \label{BCT}
\ee
\end{subequations}

The periodic standing waves are characterised by their energy
\be
E=  4 \pi\int_0^R\left( \frac{\phi_t^2}{2} +\frac{\phi_r^2}{2} + \phi^2-\phi^3+\frac{\phi^4}{4} \right) r^2dr
\label{E}
\ee
and
frequency
\be
\omega = \frac{2\pi}{T}.     \label{freq}
\ee 
If  the solution with frequency $\omega$ does not change appreciably as $R$ is increased
--- in particular, if the energy \eqref{E} does not change --- this standing wave provides a fairly accurate approximation for the periodic solution in an infinite space.

In what follows, we present results of numerical and asymptotic analysis of the 
 boundary-value problem \eqref{2D}.
Numerically, we employed a predictor-corrector  algorithm  with a  newtonian iteration
to continue   solutions
in $\omega$ \cite{numerical_parameters}.
To classify the 
stability of the resulting standing waves against spherically-symmetric perturbations we considered the
 linearised equation
\be
y_{tt}-y_{rr}- \frac{2}{r} y_r 
 -y+ 3  (\phi-1)^2 y=0
\label{ytt} \ee
with the boundary conditions $y_r(0,t)=y(R,t)=0$.
The solution  $\phi(r,t)$ is deemed stable if all its
 Floquet multipliers lie on the unit circle $|\zeta|=1${}
 and unstable if there are multipliers outside the circle   {}{\cite{Grimshaw,Chicone}.}
 The monotonically growing instability is associated with a pair of real multipliers,
 $\zeta$ and $1/\zeta$; the oscillatory instability is characterised by a complex 
 quadruplet: $\zeta, 1/\zeta, \zeta^*, 1/\zeta^*$. 

  The paper is organised into five sections.  In the next section 
  we establish the existence of a sequence of
 standing waves with $n-1$ nodes ($n=1,2,  ...$) and no clearly defined core.
  These Bessel-like patterns %   (or simply as  {\it Bessel waves\/} in what follows) 
  are nonlinear
  descendants of linear standing waves in the ball. 
  The subsequent asymptotic analysis (section \ref{large_ball}) focusses on the evolution of the $n=1$ Bessel wave as its frequency 
  is decreased to below the frequency of the spatially uniform oscillations. Further frequency reduction is carried out using  numerical continuation;
    the resulting resonant energy-frequency diagram is presented 
  in section \ref{resonances}. We consider the spatiotemporal structure of the resonant solutions and demonstrate
  that they are born in the period-doubling bifurcations of the $n>0$  Bessel waves. Stability of the standing waves is  classified   in the same section. 
  Finally, section \ref{conclusions} summarises  results of this study.

 \section{Birth of  the Bessel wave}
 \label{birth}

 We start our analysis by considering   {}{
the emergence of  a  standing wave
 from the zero solution  of equation (2a). }
  The  small-amplitude  standing wave
 can be constructed 
 as a power series
 \be
 \phi=\epsilon \phi_1+ \epsilon^2 \phi_2+ \epsilon^3 \phi_3 + ...,
 \label{S1}
 \ee
 where the coefficients $\phi_n$ are functions of $x$ and
 a hierarchy of time scales $\mathcal T_0=t$,
 $ \mathcal T_1= \epsilon t$, $ \mathcal T_2= \epsilon^2 t$, ... . In the limit $\epsilon \to 0$ the time scales become independent; hence
\[\frac{\partial^2}{\partial t^2}= \frac{\partial^2}{\partial \mathcal T_0^2} + 2 \epsilon \frac{\partial}{\partial  \mathcal T_0} \frac{\partial}{\partial \mathcal T_1} + 
\epsilon^2 \left( \frac{\partial^2}{\partial  \mathcal T_1^2} + 2 \frac{\partial}{\partial  \mathcal T_0} \frac{\partial}{\partial \mathcal T_2} 
\right) + ...
\]
Substituting the above expansions in \eqref{A200} we set to zero coefficients of like powers of $\epsilon$. 

 The solution to the order-$\epsilon$ equation, satisfying the boundary conditions $\partial_r \phi_1(0,t)= \phi_1(R,t)=0$, is
 \be
 \label{G2} 
 \phi_1= \left( A e^{i \Omega^{(n)}  \mathcal T_0} + c.c. \right) f_1^{(n)}(r),
 \ee
 where 
     \begin{align}
 \Omega^{(n)}= \sqrt{\omega_0^2+ \left( k^{(n)}\right)^2},     \quad  \omega_0= \sqrt 2,
 \label{G16}  \\
 f_1^{(n)}= \frac{\sin (k^{(n)} r)}{r}, \quad
 k^{(n)}= \frac{\pi}{R} n, 
  \label{G160}   \end{align}
  $n=1,2, ... $, 
  and $c.c.$ stands for the complex conjugate of the immediately preceding term.  
 The amplitude $A$ is slowly changing in time: $A=A( \mathcal T_1, \mathcal T_2, ...)$.
   Since the  localised mode \eqref{G160}  has the form of the spherical Bessel function, we will be referring to  solutions 
   {}{branching off the zero solution}  at $\omega=\Omega^{(n)}$  as 
  ``Bessel waves". 

 In equation \eqref{G16}, 
$\omega_0$
  demarcates the endpoint of the continuous spectrum of frequencies  in the ball of an infinite radius.
  This endpoint defines a natural frequency scale that will regularly occur in the following analysis.

The order-$\epsilon^2$   solution, satisfying $\partial_r \phi_2 (0,t)=\phi_2(R,t)=0$,  is given by
 \be
 \label{G300}
 \phi_2=\left( 3A^2 e^{2i \Omega^{(n)}  \mathcal T_0} + c.c. \right) f_2^{(n)} (r)+ 6|A|^2 g_2^{(n)}(r),
 \ee
 where 
 \begin{align}
 f_2^{(n)}= \frac{1}{\kappa^{(n)} r} \left( p^{(n)}(r)- \frac{\sin(\kappa^{(n)}r)}{\sin( \kappa^{(n)} R)} \, p^{(n)}(R) \right),
 \nonumber \\
  g_2^{(n)}= \frac{1}{\sqrt{2} r} \left( q^{(n)}(r)- \frac{\sinh(\sqrt{2} r)}{\sinh( \sqrt{2} R)} \, q^{(n)}(R) \right),
 \nonumber \\
 p^{(n)}(r)= \int_0^r \sin \left[ \kappa^{(n)} (r'-r) \right]  \frac{\sin^2 (k^{(n)} r')}{r'} dr', \nonumber \\
  q^{(n)}(r)= \int_0^r \sinh \left[ \sqrt{2}  (r'-r) \right]  \frac{\sin^2 (k^{(n)} r')}{r'} dr'
  \end{align}
  and
 \[
 \kappa^{(n)}= \sqrt{ 6+ 4 (k^{(n)} )^2}.
 \]
The solution \eqref{G300} exists provided the amplitude satisfies the nonsecularity constraint $\partial A/ \partial  \mathcal T_1=0$. 
We are also assuming that $\kappa^{(n)}  \neq k^{(m)}$, $m=1,2, ...$.

 Finally, the order $\epsilon^3$ gives an equation for $\phi_3$:
 \be
 \label{G5} 
 \left( \frac{\partial^2}{\partial  \mathcal T_0^2} -\nabla^2+2 \right) \phi_3= -2  \frac{\partial^2 \phi_1}{\partial  \mathcal T_0  \partial \mathcal  T_2}  
 +6 \phi_1  \phi_2 -\phi_1^3.
 \ee
  The solvability condition is 
 \be
% \label{G16}
  \label{S16}
   i \Omega^{(n)} R \frac{\partial A}{\partial \mathcal T_2} + 3 \sigma^{(n)} |A|^2A=0,
 \ee
 where
 \be
 %\label{G7}
 \label{S7}
 \sigma^{(n)}= \int_0^R    \left[ (f_1^{(n)} )^2 -12 g_2^{(n)} - 6 f_2^{(n)} \right]    \left( f_1^{(n)}  r\right)^2    dr,
 \ee
 and we have used   $\partial A/ \partial  \mathcal T_1=0$.

   \begin{figure}[t]
 \begin{center} 
     \includegraphics*[width=\linewidth] {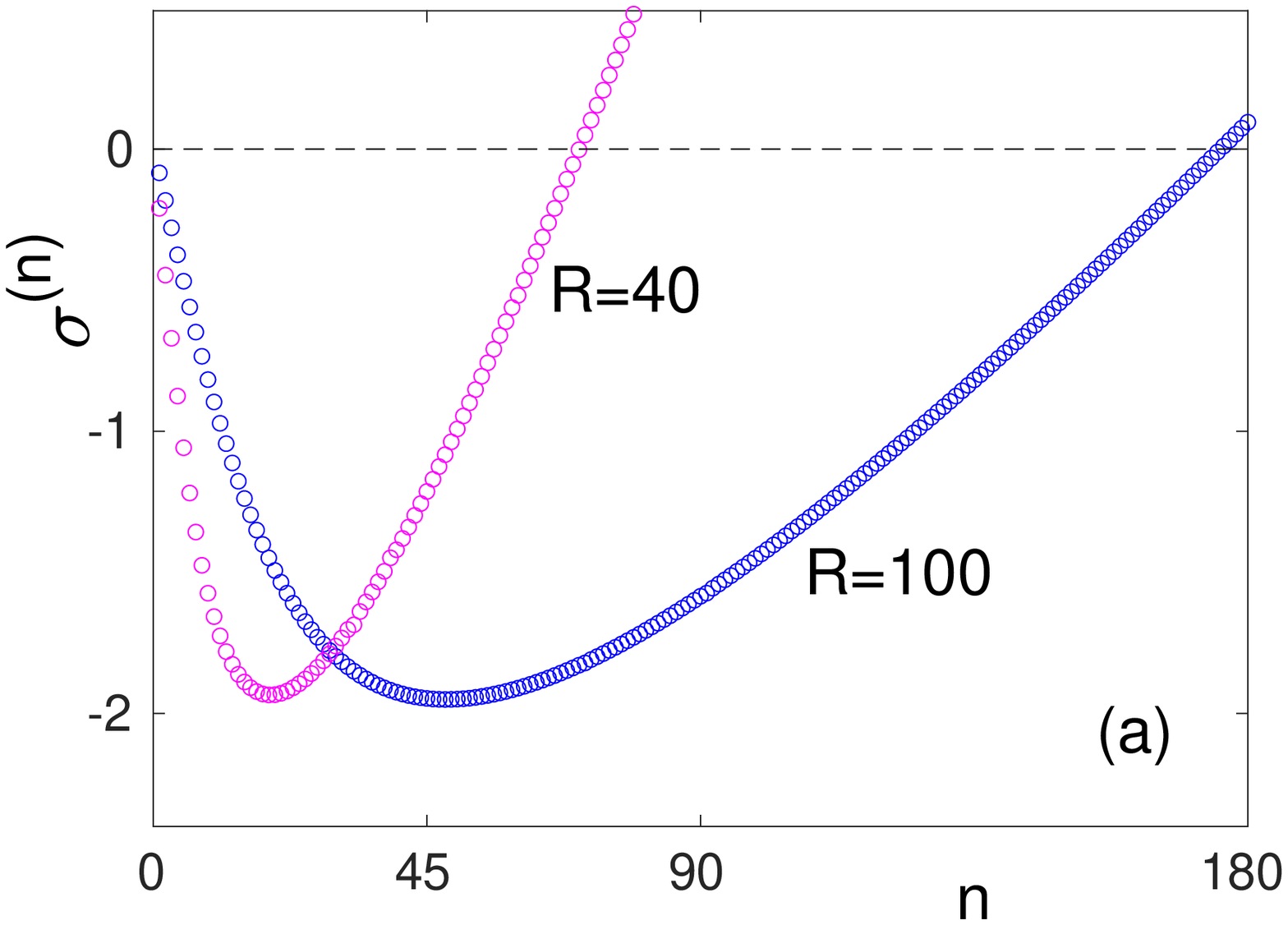}     
     
     \vspace*{4mm}
                       \includegraphics*[width=\linewidth] {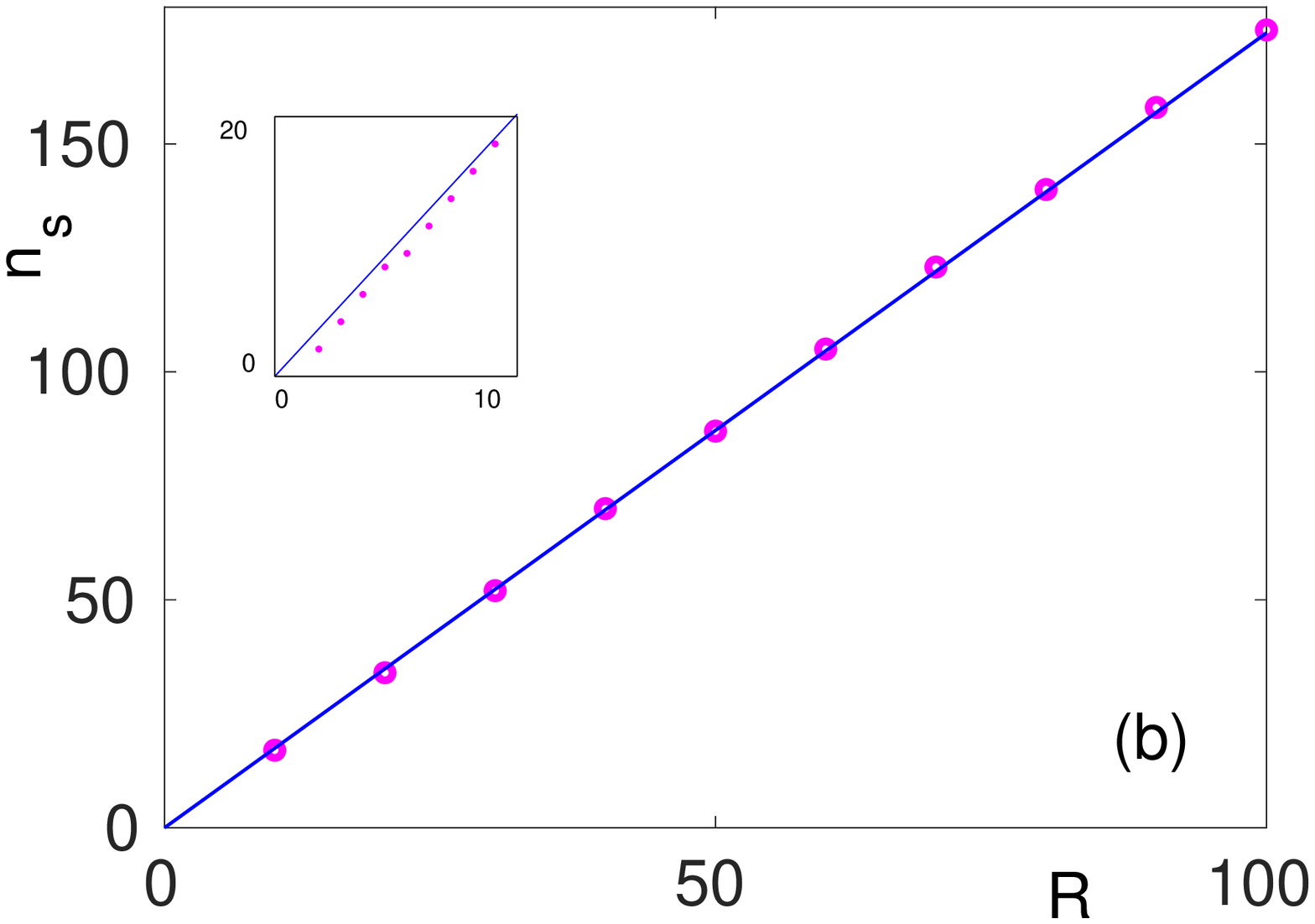}  
                   \end{center}
  \caption{(a) The integral \eqref{S7} for $R=40$ and $R=100$. The function 
  $\sigma^{(n)}$ is negative for all $n \leq n_s(R)$ and positive for $n >n_s(R)$. 
(b) The  integer $n_s$ (marked by circles)  for a sequence of $R$ values. For  $R\geq 2$, the function $n_s(R)$ is well approximated by $n_s=(\sqrt{30}/ \pi) R$
(shown by the straight line). The inset blows up the interval $2 \leq R < 10$. 
\label{sigma_n} 
}
 \end{figure}  
 
 The values of the  integral \eqref{S7} with varied $n$ are presented graphically in Fig \ref{sigma_n}. 
 The quantity   $\sigma^{(n)}$ is determined to be negative for all $n  \leq n_s$
 and positive for $n>n_s$, where $n_s$ is an integer dependent on $R$. When $R$ is large enough, a fairly accurate approximation for
 $n_s(R)$ is given by the integer part of  $(\sqrt{30} / \pi)  R$.

 The general solution of the amplitude equation \eqref{S16} is
 \be
 A= \exp \left( i  \frac{3 \sigma^{(n)}}{\Omega^{(n)} R} \mathcal  T_2 \right),
  \label{G8}
 \ee
 where the initial value  was set equal to 1.
 (There is no loss of generality in setting $A(0)$ to 1  as it enters $\phi_n$ only in combination $ \epsilon A(0)$,
 where $\epsilon$ is free to vary.) 
 Thus, the fundamental frequency of the Bessel wave with amplitude $\epsilon$, {}{branching off the trivial solution  $\phi=0$ at the point $\omega=\Omega^{(n)}$,}   is
 \be
 %\label{G10} 
 \label{S10} 
 \omega= \Omega^{(n)}+  
  \frac{3 \sigma^{(n)}}{ \Omega^{(n)}R}
  \epsilon^2 + ... .
 \ee 
 Note that the nonlinear frequency shift 
  is negative ($\omega< \Omega^{(n)}$) 
   for all $n  \leq n_s$ 
  and positive for $n > n_s$.

  The relation \eqref{S10} implies that our $\epsilon$-expansion is, in fact, 
  an expansion in powers of the detuning from the resonant frequency,
  $|\omega- \Omega^{(n)}|$.

 The energy \eqref{E} of the series solution \eqref{S1} is
 \be
 \label{E_pert}
 E^{(n)}(\epsilon) = 4 \pi R \left( \Omega^{(n)} \right)^2  \epsilon^2 + O(\epsilon^4).
 \ee
 % $E^{(n)}(\epsilon) >0$. 
Eliminating $\epsilon^2$ between \eqref{S10} and \eqref{E_pert} we can express the energy of the Bessel wave as a function of its frequency: 
\be
E^{(n)}(\omega)   = \frac{ 4 \pi R^2 \left( \Omega^{(n)}\right)^3 }{3 \sigma^{(n)}} \left(\omega-\Omega^{(n)}\right) +O\left( \left(\omega-\Omega^{(n)}\right)^2\right). 
\label{S12}
\ee 
This is an equation of a ray emanating from the point $(\Omega^{(n)}, 0)$ on the $(\omega, E)$, $E>0$,
 half-plane. 
The slope of 
 the ray is negative  for all $n \leq n_s(R)$ and positive for $n >n_s(R)$.

All solutions of equation \eqref{S16} are stable. (Trajectories form concentric circles on the $(\mathrm{Re}  A, \mathrm{Im} A)$ phase plane.)

\begin{figure}[t]   \vspace*{-4mm}
 \begin{center}  \hspace*{-7mm}
              \includegraphics*[width=1.1 \linewidth] {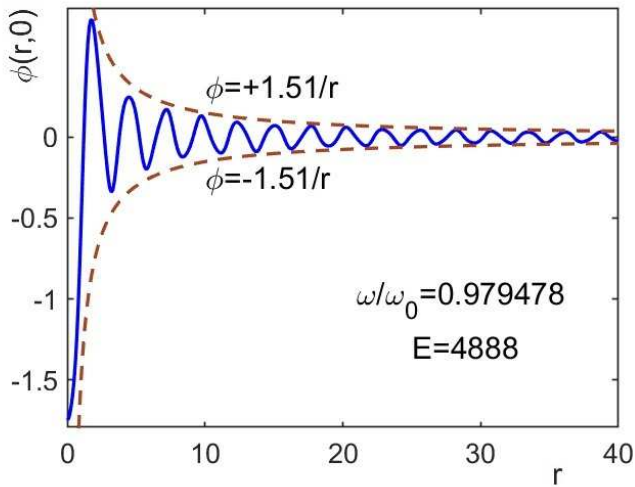} 
                                           \end{center}
                                           \vspace*{-2mm}
  \caption{\label{Bess} 
 A snapshot of the Bessel wave with high  energy. % (Here $E=4888$.) 
   This solution was obtained by the numerical continuation of 
  the trivial solution from $\omega= \Omega^{(114)}$, in a ball with
 $R=150$. (For the entire Bessel branch see Fig \ref{doubling}.)
  The wave is depicted by a solid line while 
   its dashed  envelope highlights the absence 
  of a well-defined core. Note that only a portion of the $(0,150)$ interval is shown.
    }
 \end{figure}

The asymptotic construction of the Bessel wave is corroborated by the
 numerical  analysis of  the boundary-value problem \eqref{2D}.
A  numerically-continued branch starting with  the trivial solution $\phi=0$ 
 at $\omega= \Omega^{(n)}$ consists of 
 standing waves with $n-1$ nodes inside the interval $(0,R)$.
An important feature of these solutions 
is their  weak localisation.  Even when the energy of the Bessel wave  is high
--- that is, even when the solution is far from its linear limit \eqref{G160}  ---  the 
wave does not have an exponentially localised core and the
 amplitude of the damped sinusoid  $\phi(r,t)$ remains of order  $R^{-1}$
 as $r$ approaches $R$. (See 
Fig \ref{Bess}.)

Consistently with the asymptotic considerations, the numerically continued  Bessel waves 
are stable near their inception points and only
lose stability as their energies become high enough.
(For details of the corresponding period-doubling 
bifurcation see section \ref{4a}.)

\begin{figure}[t]
 \begin{center} 
            \includegraphics*[width=\linewidth] {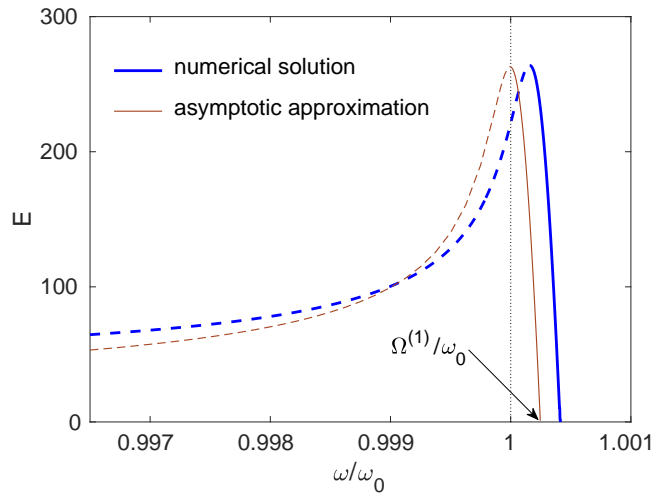} 
              \end{center}
  \caption{
  The  $E(\omega)$ dependence near the point of inception of the standing wave
   in a ball with  $R=100$.
   Blue (thick) curve:  result  of numerical continuation. 
  Brown (thin) curve: asymptotic approximation exploiting $R^{-1}$ as a small parameter.
  Stable solutions are marked by the solid and unstable ones by the dashed lines. 
  \label{near1}
  }
 \end{figure}

The continuation 
starting at the lowest of the resonance values,
$\omega=\Omega^{(1)}$,  produces  a stable branch with a steep negative slope (Fig \ref{near1}).
The steep growth of the energy is due to the small absolute value of $\sigma^{(1)}$  in \eqref{S12}
while the negativity of $dE^{(1)}/d\omega$  is due to $n_s(100)$ being greater than 1.
As the solution is continued to lower values of $\omega$,
 the function $E(\omega)$ 
 reaches a maximum and starts decreasing. 
 Not unexpectedly, the asymptotic expansion in powers of the small detuning $|\omega-\Omega^{(1)}|$
  does not capture the formation of the energy peak.

  Before turning to an asymptotic expansion about a different frequency value, we make 
   a remark on the nomenclature of numerical solutions. 
 Assume that the 
 computation interval $(0,  T)$ includes an integer number of fundamental periods of a  
  solution  of the boundary-value problem \eqref{2D}:
 $ T= m  T_{\mathrm{f}}$, $m>1$. 
 Equation \eqref{freq} gives then a formal frequency $\omega=  \omega_{\mathrm f}/m$, where $  \omega_{\mathrm{f}}= 2 \pi/ T_{\mathrm{f}}$
 is  the fundamental frequency of the wave. In this case the  periodic solution $\phi(r,t)$
 will be referred to as the $1/m$  undertone of the standing wave.

 It is important to emphasise that the only difference between a standing
 wave and its undertone is the length of the  interval   $(0,T)$ that we use to determine the respective solution ---
 and hence  its formal frequency \eqref{freq}. For example, the $n$-th Bessel wave is born with the frequency $\omega= \Omega^{(n)}$ 
 while its $1/2$ undertone is born with $\omega= \Omega^{(n)}/2$.
 Basically, the $1/2$ undertone of the periodic oscillation $\phi(r,t)$  is the oscillation itself, where we skip
 every other beat.

 \section{Small-amplitude  wave in a large ball}
 \label{large_ball}
 
 \subsection{Inverse radius as a small parameter} 
 
 In order to account for the energy peak in Fig \ref{near1} and
 track the $E(\omega)$ curve over the point of maximum analytically, we  need an asymptotic expansion of a
 different kind. Instead of assuming the proximity to the resonant frequency $\omega=\Omega^{(1)}$,
   we will zoom in on the neighbourhood of the frequency $\omega_0$ corresponding to the
 uniform oscillations in an infinitely large ball. Our approach is a relative 
 of the Lindstedt-Poincare method utilised in the context of the  infinite space in
 Ref \cite{Voronov2,B} and elucidated in \cite{Fodor2}.
   (The method was pioneered  in the one-dimensional setting  \cite{Kose,Dashen}.)

 We construct the small-amplitude solution in a ball of a large --- yet finite --- radius. 
 Instead of  the techniques used  in \cite{Voronov2,B,Fodor2,Kose,Dashen}, we employ
 a multiple scale expansion.
This approach affords information on the spectrum of
 small perturbations  of the standing wave, in addition to the standing wave itself.

The inverse radius $\epsilon=R^{-1}$ provides a natural small parameter.
We expand $\phi$ as in \eqref{S1}, introduce the sequence of slow times $\mathcal T_n$ and, in addition, 
define a hierarchy of spatial scales ${\bf X}_n=\epsilon^n {\bf x}$.
% \be   \label{E30}
% \phi= \epsilon \phi_1 +  \epsilon^2  \phi_2 +  \epsilon^3 \phi_3+ ...  \ .
% \ee
Hence
\[
\nabla =  \nabla_0 + \epsilon \nabla_1 + \epsilon^2 \nabla_2 + ...  ,
\quad \nabla_n = \frac{\partial}{\partial {\bf X}_n}.
  \]
These expansions are substituted in the equation \eqref{A200} where,
for ease of computation,  we drop the requirement of spherical symmetry:
\be
\phi_{tt}- \nabla^2 \phi + 2 \phi- 3 \phi^2 + \phi^3=0.
\label{A2000}
\ee

At the order $\epsilon^1$,
 we choose a spatially homogeneous solution 
 \be
 \label{E2}
 \phi_1= A e^{i \omega_0 \mathcal T_0} + c.c.
 \ee
In \eqref{E2},  the amplitude  $A$ does not depend on ${\bf X}_0$ or $ \mathcal T_0$ but may depend on the ``slower" variables
${\bf X}_1, {\bf X}_2, ...$ and $\mathcal T_1,  \mathcal T_2, ...$. 
% The letters $c.c.$ stand for the complex conjugate of the   immediately preceding term.
 
 The order $\epsilon^2$ gives 
 \be
 \label{E7}
 \phi_2= 3|A|^2 -\frac12 A^2 e^{2 i \omega_0  \mathcal T_0} + c.c.,
 \ee
 and we had to impose the constraint $\partial A/ \partial  \mathcal T_1=0$. 
  Proceeding to the cubic order in $\epsilon$ we obtain 
 \begin{align}
  \left(   \frac{\partial^2}{\partial  \mathcal T_0^2}  - \nabla_0^2 +2 \right) \phi_3=  - 4 A^3 e^{3 i \omega_0 \mathcal T_0} + c.c.  \nonumber \\ 
 + \left(  \nabla_1^2 A -2i \omega_0 \frac{\partial A}{\partial  \mathcal T_2} 
   +12 |A|^2A    \right) e^{i \omega_0  \mathcal T_0}   + c.c.
  \label{E9}
 \end{align}
Setting to zero the secular term in the second line of \eqref{E9}, we arrive at the 
amplitude equation 
\begin{subequations}  \label{ampl}
\be
-2i \omega_0 \frac{\partial A}{\partial  \mathcal T_2}  +
\nabla_1^2 A 
   +12 |A|^2A=0.
   \label{E10}
   \ee  
The boundary condition $\phi(R,t)=0$  translates into 
   \be
   \left. A({\bf X}_1, \mathcal  T_2)   \right|_{|{ \bf X}_1|=1} =0.
   \label{F2} 
   \ee
   \end{subequations} 
 
 \subsection{Schr\"odinger equation in a finite ball} 
 \label{ball}

\begin{figure}[t]
 \begin{center} 
    \includegraphics*[width=\linewidth] {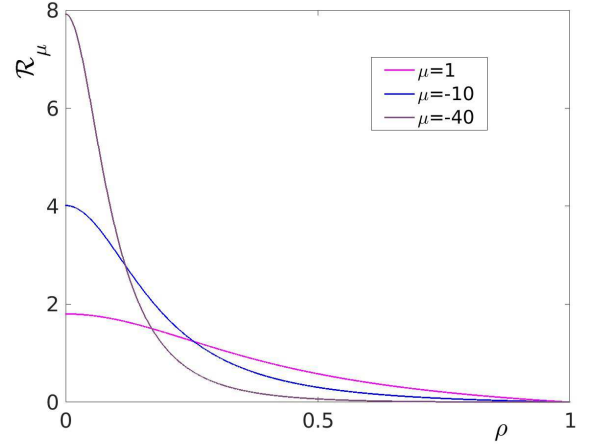}      
  \end{center}
  \caption{$\mathcal R_\mu(\rho)$:
  the nodeless solution of the boundary-value problem \eqref{E14}.
  As $\mu$ changes from negative to positive values, the exponentially localised 
  solution gives way to a function without a clearly defined core.
  \label{Rcal}
  }
 \end{figure}

 A family of  spherically-symmetric solutions of \eqref{ampl} is given by
 \be
 \label{E15}
 A=      e^{i  \omega_2  \mathcal T_2}    \mathcal R_\mu (r_1),
 \ee
 where
 $r_1= \sqrt{ {\bf X}_1^2}$ and
 $\mathcal R_\mu(\rho)$ solves the  boundary-value problem
 \begin{subequations} 
  \label{E14}
 \begin{align}
 \mathcal R^{\prime \prime} + \frac{2}{\rho} \mathcal R^\prime + \mu \mathcal R + 12 \mathcal R^3=0,
 \label{E11} \\
 \mathcal R^\prime(0) = \mathcal R(1) = 0,
 \label{E12}
 \end{align}
 \end{subequations}
 with $\mu= 2 \omega_0 \omega_2$. 
 (In \eqref{E14}, the prime stands for the derivative with respect to $\rho$.) 
  In what follows we
 confine our attention  to  the nodeless (everywhere positive) solution  $\mathcal R_\mu(\rho)$
 (Fig \ref{Rcal}). 
 Of particular importance will be its norm squared,
 \be
 \label{N1}
  N(\mu) = \int_0^1   \mathcal R_\mu^2 (\rho) \rho^2 d \rho.
  \ee

 The nodeless solution $\mathcal R_\mu (\rho)$ exists for all $\mu$ with $-\infty< \mu < \pi^2$. 
 As $\mu \to \pi^2$, a  perturbation argument gives 
 \begin{subequations}  \label{J1} 
 \begin{align}
 \mathcal R_\mu(\rho)  = \alpha  \sqrt{\pi^2-\mu}  \,  \frac{ \sin( \pi \rho)}{\rho} + O\left(  (\pi^2-\mu)^{\frac32}   \right),         \\
 \alpha^2= \frac{1}{12 \pi} \frac{1}{2 \mathrm{Si} (2 \pi) - \mathrm{Si} (4 \pi)}= 1.973 \times 10^{-2},  
 \end{align}
 \end{subequations}
 so that  the norm decays to zero:
 \[
 N(\mu)= \frac{\alpha^2}{2}( \pi^2-\mu)+ O\left(\left(\pi^2-\mu \right)^2 \right).
 \]
 As $\mu \to -\infty$, we have 
 \be
 \mathcal R_\mu (\rho) \to  \sqrt{ - \mu} \, S\left(\sqrt{-\mu} \rho \right),
 \label{scaling} 
 \ee
 where  $S(\rho)$ is the nodeless solution 
 of the boundary value problem
 \begin{subequations} \label{Sbvp}
 \begin{align}
 S^{\prime \prime}  + \frac{2}{\rho} S^\prime - S + 12S^3=0, \\
 S^\prime (0)= S(\infty)=0.
 \end{align}  \end{subequations}
 Accordingly, the norm \eqref{N1}  decays to zero in the latter limit as well:
 \[
 N(\mu) \to \frac{1}{ \sqrt{-\mu}} \int_0^\infty S^2(\rho) \rho^2 d \rho
 \quad \mbox{as}  \ \mu \to -\infty.
 \]
 The numerical analysis of the problem \eqref{E14} verifies that $N(\mu)$  has a single maximum, at $\mu_c= -0.225$. 

 Thus we have constructed an asymptotic standing-wave solution of equation \eqref{2D}, parametrised by its frequency $\omega$:
 \begin{subequations} 
 \label{E16}
 \be
 \phi=   \frac{2}{R}
 \cos (\omega  t ) \mathcal R_\mu \left(\frac{r}{R}  \right) +O(R^{-2}),
 \label{phiR}
 \ee
 where 
 \be
\mu= 2 \omega_0R^2 (\omega-\omega_0).
\label{muom}
\ee
\end{subequations}
Substituting \eqref{phiR} in \eqref{E} we obtain 
  the corresponding energy:
  \be
  \label{Eas}
  E(\omega) =   16 \pi  R N(\mu)+ O(R^{-1}).
  \ee
 
   The dependence \eqref{Eas} is shown by the thin line
   in Fig \ref{near1}. Unlike the expansion  in powers of the frequency detuning  (section \ref{birth}), 
   the expansion in powers of $R^{-1}$ is seen to reproduce the energy peak.
   The 
    peak of the curve $E(\omega)$ is a scaled version of the peak of 
    $N(\mu)$.

  Finally, we note that the function $\mathcal R_\mu(\rho)$ with  negative $\mu$ has an
 exponentially localised core, with the width of the order  $\frac{1}{\sqrt{-\mu}}$. 
 By contrast, solutions with $\mu>0$ approach zero at a nearly uniform rate (Fig \ref{Rcal}).

  \subsection{Stability of small-amplitude standing wave} 
  \label{sta_small} 
 
 By deriving the amplitude equation \eqref{ampl} the analysis of stability of the time-periodic standing wave
 has been  reduced to the stability problem for the stationary solution of the 
 3D nonlinear Schr\"odinger equation. The leading order of a linear perturbation to
 the solution  \eqref{E16} is given by 
 \begin{align}
 \delta \phi=  \left\{
  e^{i  \omega_0  \left(1
 + \frac{\mu}{4R^2}  \right)t }
  \left[  \mathcal{F} \left(  \frac{r}{R}   \right)  +i  \mathcal{G} \left(  \frac{r}{R} \right)                      \right]           +   c.c. \right\}        \nonumber \\ 
   \times
 \exp \left( \frac{\lambda}{2 \omega_0 R^2 }   t  \right),
 \label{E50} 
 \end{align}
 where $\mathcal{F}=\mathcal{F} (\rho)$ and $\mathcal{G} =\mathcal{G} (\rho)$ are two components of an eigenvector
 of the symplectic   eigenvalue problem
\be  \label{E34} 
 L_0 \mathcal{G} = - \lambda \mathcal{F}, \quad L_1 \mathcal{F} = \lambda \mathcal{G}.
 \ee
 In \eqref{E34}, $L_0$ and $L_1$ are a pair of  radial operators
 \begin{align}
L_0= - \frac{d^2}{d \rho^2} - \frac{2}{\rho} \frac{d}{d \rho} -\mu  - 12 \mathcal R_\mu^2(\rho),     \nonumber 
  \\
L_1= - \frac{d^2}{d \rho^2} - \frac{2}{\rho} \frac{d}{d \rho}   -\mu  - 36 \mathcal R_\mu^2(\rho),
\label{L0L1} 
\end{align}
with the boundary conditions 
\be
\mathcal{F}^\prime(0)=\mathcal{G}^\prime(0)= \mathcal{F}(1)=  \mathcal{G} (1)=0.
\label{F6}
\ee

The lowest eigenvalue of the Schr\"odinger operator $L_0$ is zero, with the associated eigenfunction given by $\mathcal R_\mu(\rho)$.
Numerical analysis reveals that the operator $L_1$ has a single negative eigenvalue.
% while zero is not an eigenvalue of $L_1$.
 This is the case of applicability of the Vakhitov-Kolokolov criterion  \cite{VK,Peli1,Peli2}. 
 The criterion guarantees the stability  of the solution \eqref{E15}  if $dN/ d \mu<0$ and instability otherwise.

Numerical methods confirm that in the region  $\mu_c<\mu<\pi^2$,
 the eigenvalue problem \eqref{E34}-\eqref{F6} does not have any real eigenvalues apart from a pair of zeros resulting from the U(1) invariance of \eqref{E10}. 
 (We remind that $\mu_c$ is the point of maximum of the curve $N(\mu)$;  $\mu_c=-0.225$.) 
 As $\mu$ is decreased through $\mu_c$, 
  a   pair of opposite pure-imaginary   eigenvalues $\pm \lambda_0(\mu)$ converges at the origin and diverges along the positive and negative real axis.
  As $\mu \to -\infty$, the scaling \eqref{scaling} gives $\lambda_0(\mu) \to -5.50 \mu$, where $5.50$ is the symplectic eigenvalue associated with the 
  solution of the 
  infinite domain problem \eqref{Sbvp}.

  The upshot of our asymptotic analysis is that there is a continuous family of standing-wave solutions in the ball of a  large radius $R$,
  with frequencies $\omega$  extending down from $\Omega^{(1)}$.  The function $E(\omega)$ features a sharp peak  
  at $\omega_c = \omega_0 + \mu_c (2 \omega_0 R^2)^{-1}$,
  with the
  standing waves to the right of the peak (where $dE/ d\omega<0$) being stable and those to the left  (where $dE/ d\omega>0$) unstable. 
  (See the thin curve in Fig \ref{near1}.)

\subsection{Continuation over the energy peak} 
\label{overpeak}

          \begin{figure}[t]
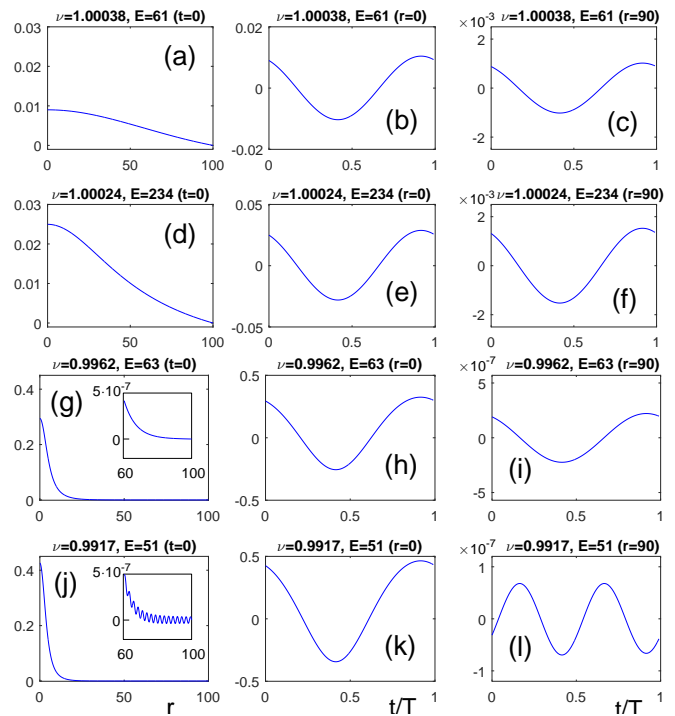

 \begin{center} 
       \includegraphics*[width=\linewidth] {fig_5a.eps}  
  \includegraphics*[width=\linewidth] {fig_5b.eps}   
  \end{center}
  \caption{Top to bottom row: standing wave as it is continued from $\Omega^{(1)}$ to lower $\omega$ in  Fig \ref{near1}. 
  (The ball radius $R=100$.)
  Left column: spatial profile  at  a particular time, $\phi(r,0)$. Middle and right column:  temporal behaviour at the central and a peripheral 
  point, $\phi(0,t)$ and $\phi(90,t)$. In the panel legends,  $\nu$ is the normalised frequency: $\nu= \omega/ \omega_0$.
  The same notation is used in Figs \ref{6r}, \ref{2nd_Bessel}, \ref{left_curve}
  \label{12panel}
  }
 \end{figure}  

The large-$R$ perturbation expansion with $\omega$ close to
 $\omega_0$
 was validated by  the numerical study  of the  boundary-value problem \eqref{2D}.
 We continued the periodic solution $\phi(r,t)$ to lower $\omega$ 
 and used the linearised equation \eqref{ytt} to evaluate the associated monodromy matrix.
 % For each numerical $\phi(r,t)$ we used the linearised equation \eqref{ytt} to evaluate the associated monodromy matrix.
  In agreement with the asymptotic considerations, 
  a pair of real  Floquet multipliers ($\zeta$ and $\zeta^{-1}$) 
  was seen to leave  the unit circle as
 $\omega$ passed through the point of maximum energy. Consequently, the left slope of the energy peak in Fig \ref{near1} does indeed
  correspond to unstable standing waves.

Fig.\ref{12panel} documents  the  solution 
as it is  continued from $\Omega^{(1)}$ over the energy peak.
Consistently with the asymptotic expression  \eqref{E16}, the peripheral field values $\phi(r_p,t)$, where $r_p \sim R$,  oscillate at the same frequency  $\omega$ as the amplitude at the origin,
 $\phi(0,t)$. This agreement is recorded on either side of the energy peak;
 see panel pairs (b) and (c), (e) and (f), (h) and (i).
 
  As $\omega$ is reduced below the point of maximum  energy, 
    the Bessel-function profile  \eqref{E16}, \eqref{J1}  gives way to an exponentially localised shape.
     This metamorphosis   agrees   with the evolution of the asymptotic profile $\mathcal R_\mu(\rho)$ as $\mu$ is taken from positive to negative values.
     The difference in the type of decay is clearly visible in 
   panels (a), (d)  and (g) of     Fig  \ref{12panel}. 
   Lowering $\omega$ even further sees the formation of a 
   small-amplitude undulating  tail (Fig  \ref{12panel}(j)).
  At the same time, 
   the oscillation frequency in the peripheral region 
 switches from the  frequency of the core of the standing wave to its
 second harmonic 
 (compare panel (l) to (k)).

    It may seem that the presence of the 
     second-harmonic  tail is at variance with the uniformly-first  harmonic pattern \eqref{E16}. There is no contradiction, in fact. 
 As we take $\omega$ far enough from $\omega_0$, 
    the assumption $\phi =O(R^{-1})$ becomes
 invalid and the expression \eqref{E16} stops providing any accurate approximation to the solution $\phi(r,t)$.

Why does the formation of the second-harmonic tail require taking  $\omega$ far from $\omega_0$? The reason is that 
when $\omega$ is close to $\omega_0$,  the core of the exponentially-localised  standing wave  is much wider than 
the wavelength of the second-harmonic radiation: 
\[
\frac{1}{\sqrt{
2 \omega_0 (\omega_0-\omega)}} \gg   \frac{2 \pi}{\sqrt {3 \omega_0^2}}.
 \]
(Here we took advantage of the fact the characteristic width of the bell-shaped function $\mathcal R_\mu(\rho)$  is $1/ \sqrt{-\mu}$ 
and used \eqref{muom} to express $\mu$.)
 As a result, the radiation coupling to the 
core is weak and its amplitude is exponentially small.   Thus when $\omega$ is close to $\omega_0$,   we can simply not discern the amplitude of the
second harmonic against the first-harmonic oscillation.

  \subsection{Small-amplitude wave in the infinite space}

    {}{
  It is instructive to comment on the $R \to \infty$
   limit  
  for which the small-amplitude solution   is available in  the earlier literature    \cite{Voronov2,B,Fodor2}.}
  
  In the case 
    of the infinitely large ball 
   our asymptotic expansion remains in place
but  $\epsilon$ becomes a formal expansion parameter,  not tied to $R$. Without loss of generality, we can let $\mu=-1$ in 
 equation \eqref{E11} while  the boundary condition $\mathcal R(1)=0$ 
should be replaced with  $\mathcal R(\infty)=0$. 
{}{
In agreement with     \cite{Voronov2,B,Fodor2},
the asymptotic solution \eqref{S1} acquires the form}
\be
\phi= 2\epsilon  \cos \left[  \omega_0 \left(   1- \frac{\epsilon^2}{4} \right) t  \right] S(\epsilon r)+O(\epsilon^2),
\label{E52}
 \ee
where $\mathcal S (\rho)$ is a nodeless solution of the boundary value problem \eqref{Sbvp}.
{}{
(For solutions of  \eqref{Sbvp} see \cite{Anderson,Fodor2}.) }
As $\omega \to \omega_0$ (i.e. as $\epsilon \to 0$), the energy of the asymptotic solution \eqref{E52} tends to infinity:
\be 
E = \frac{16 \pi}{\epsilon} \int_0^\infty S^2(\rho)  \rho^2 d \rho =  \frac{16 \pi }{\epsilon}\times 0.1253.
\ee

Stability or instability of the infinite-space solution is decided by eigenvalues of the symplectic eigenvalue problem \eqref{E34}-\eqref{L0L1} with $\mu$  set to $-1$, 
$\mathcal R_\mu(\rho)$
 replaced with $S(\rho)$, and the 
boundary conditions \eqref{F6} substituted with $\mathcal{F} (\infty)= \mathcal{G} (\infty)=0$. 
The numerical analysis  verifies that the resulting symplectic problem 
has a  (single)  pair of opposite real eigenvalues $\lambda = \pm 5.50$.
Hence the solution \eqref{E52} is unstable for any sufficiently small $\epsilon$.

  \section{Resonances in the ball}
  \label{resonances}

   \begin{figure}[t]
 \begin{center} 
           \includegraphics*[width=\linewidth] {fig_6a.eps}
               
          \vspace*{3mm}
            \includegraphics*[width=\linewidth] {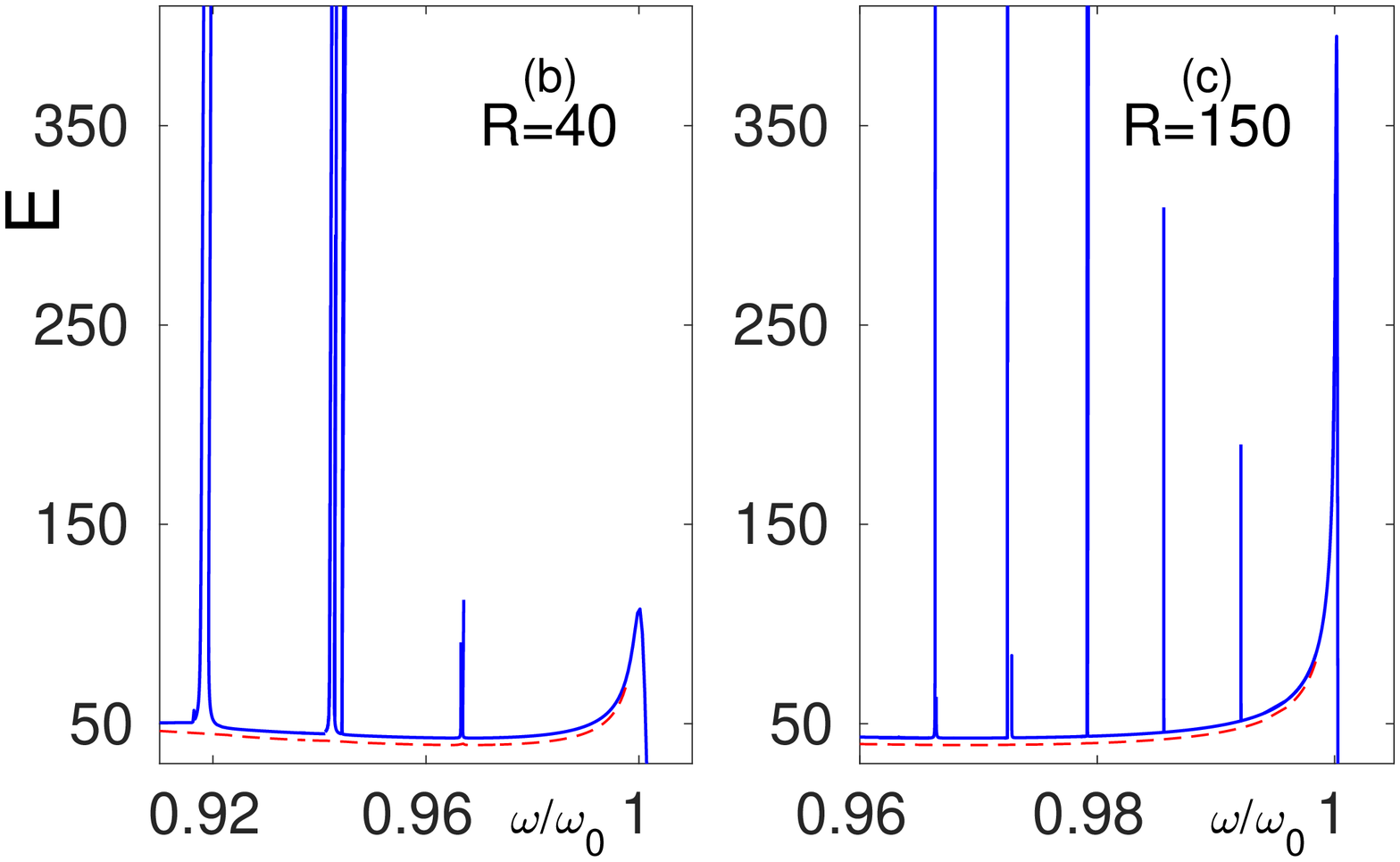}  
  \end{center}
  \caption{\label{Ew}
  (a) Energy of the standing wave with frequency $\omega$ in the ball  of  radius $R=100$ (a), 
  $R=40$ (b) and $R=150$ (c).
The $E(\omega)$ curve features a sequence of
  sharp  spikes that have complex fine structure indiscernible in the figure.
  Although some branches were only continued to moderate  energies, we expect all  spikes to extend to the top of the panels. 
  The vertical dashed lines in (a) mark the points $\omega=  \Omega^{(n)}/2$ where $\Omega^{(n)}$ are the frequencies of the newborn Bessel waves
  (defined by equation \eqref{G16}). 
The fraction next to a spike indicates the order of the Bessel-wave undertone that 
  this spike's slopes approach (but not necessarily  join) at a larger $E$.
    (The Bessel  undertones are not shown in the figure.)
 In all three panels, 
 the red dashed arc underlying the $E(\omega)$ curve is the envelope of the family of spikes. 
 For visual clarity,  it has been shifted down by a tiny amount from its actual position.
  }
 \end{figure}

  \subsection{Energy-frequency    diagram}
  \label{4a}

The numerical continuation beyond the peak in Fig \ref{near1}, from right to left,  produces an  $E(\omega)$ curve 
with what looks like a  sequence of spikes. 
Fig \ref{Ew}(a) depicts this curve for $R=100$. It also shows an envelope of the family of spikes
--- a U-shaped arc that coincides with the $E(\omega)$ curve everywhere except the neighbourhoods of the spikes. 
In the neighbourhood of each spike,  the envelope 
 bounds  it from below.

Figs \ref{Ew}(b) and (c)  compare the density of spikes in the diagrams with different values of $R$.  
(Either panel focusses on the right end of the respective diagram where spikes are thin and nonoverlapping.)
 The number and positions of the spikes are seen to be $R$-sensitive. 
  In contrast,  the U-shaped envelope
does not change appreciably as the radius of the ball
 is varied. Regardless of $R$, the U-shaped curve has a single minimum, 
 at 
 \be 
 \omega_{\mathrm{min}}=0.967 \, \omega_0.
 \label{omega_min}
 \ee

The U-shaped envelope agrees with the energy curve of  periodic infinite-space 
 solutions 
with exponentially localised cores and small-amplitude tails decaying slowly as $r \to \infty$
 \cite{Fodor1}. 
The energy of those nanopterons is defined as the integral \eqref{E} where $R$ is a radius of the core.
 The  nanopteron's  energy has a minimum at $\omega = 0.9652  \,\omega_0 $  \cite{Fodor1}
which is close to our $\omega_{\mathrm{min}}$ in \eqref{omega_min}.

 A sequence of vertical dashed lines drawn at $\omega=\frac12  \Omega^{(n)}$ in Fig \ref{Ew}(a)
  is seen to match the  sequence of spikes.
     The correspondence between the two sequences 
suggests some relation between  the spikes   and  the 
Bessel waves  born  at  $\omega=\Omega^{(n)}$.

 \subsection{Bifurcation unpacked}

\begin{figure}[t] 
 \begin{center} 
       \includegraphics*[width=\linewidth] {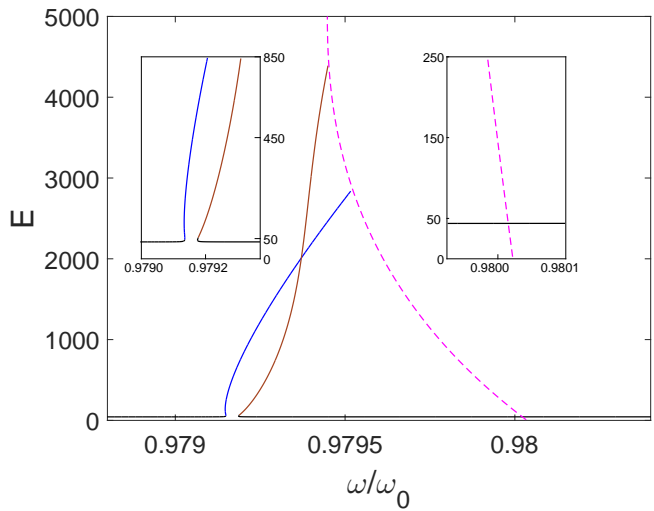}        
            \end{center} \vspace*{-3mm}
  \caption{\label{doubling}   
  A fragment of the $E( \omega)$ diagram in the vicinity of a 1:2 resonance
  in the ball with $R=150$. The blue and brown curves are two slopes of the ``spike".
  The dashed magenta arc emerging from $E=0$ at $\omega= \Omega^{(114)}/2$
   is the $1/2$ undertone of the $n=114$-th
    Bessel wave, (That is, a point $(\omega, E)$  on this branch represents the Bessel wave  with  frequency $2 \omega$.)
       The  insets zooming in on the lower sections of the  ``spike" and Bessel branch
  aim to emphasise the difference in the origins of the two branches.
  }
 \end{figure}

 Zooming in on one of the  distinctly separate spikes near the right end of the diagram
 reveals that it is not a mere peak, or projection,
 on  the $E(\omega)$ curve. As in a proper peak, there are  two energy branches 
  that rise steeply from the U-shaped arc   but instead of joining together,  the left and right ``slopes" connect to 
 another curve.  This curve turns out to be  a  Bessel branch ---
 more precisely, the $1/2$ undertone of the Bessel branch {}{emerging from  $\phi=0$  at the frequency $\omega=    \Omega^{(n)} /2$}  with some large $n$ (Fig \ref{doubling}).

\begin{figure}[t]
 \begin{center} 
    \includegraphics*[width=\linewidth] {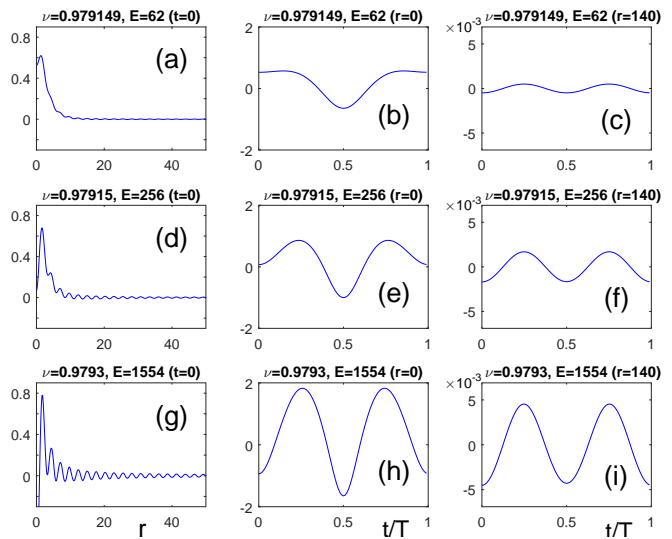}  
  \end{center}
  \caption{\label{6r} 
  Top to bottom row: spatial and temporal behaviour of the solution to the problem \eqref{2D}
   as it is continued along the blue slope of the ``spike" in Fig \ref{doubling}, from
  the underlying U-arc towards the Bessel curve.
  Left column: snapshot of $\phi(r,t)$ at a particular moment of time ($t=0$); central column: behaviour of the central value $\phi(0,t)$; right column: evolution 
  of an asymptotic value $\phi(r_p,t)$ with $r_p=140$.
All solutions satisfy the boundary condition $\phi(150,t)=0$. 
  }
 \end{figure}   
 
  To 
appreciate details of the bifurcation, 
we follow
   the curve corresponding to the left slope of the ``spike" 
 (the blue  curve in Fig \ref{doubling}). 
 A standing wave with $(\omega,E)$ located at
 the base of the ``spike"
  has an exponentially localised core and an oscillatory tail with the amplitude
  decaying in proportion to $r^{-1}$
 (Fig \ref{6r}(a)). 
 The   $\phi$-value at $r=0$ performs nearly-harmonic oscillations with the fundamental frequency $\omega= { 2 \pi}/{T}$
 (panel (b)) while the  tail oscillates at the  frequency $2 \omega$ (panel (c)).
    
  Moving  up the blue curve in Fig \ref{doubling}, 
    the contribution  of the second harmonic to the oscillation of the core increases (Fig \ref{6r}(e)). 
  Eventually, 
  when the  curve is  about to join  the  branch 
  of the $1/2$ Bessel undertones (shown by the dashed magenta   in Fig \ref{doubling}), 
    $\phi(0,t)$ completes  two nearly-identical cycles over the interval $T= 2\pi/\omega$ (Fig \ref{6r}(h)). The solution does not have any well-defined core (panel (g)), with the central and peripheral values oscillating at the same fundamental
  frequency $2\omega$ (panels (h) and (i)).
  This is exactly the spatio-temporal behaviour of the $1/2$ Bessel undertone.

  Note that the merger of  the blue and magenta curves in  Fig \ref{doubling}   can be seen as  the period-doubling bifurcation of the   Bessel wave.
  As we observed in section \ref{birth}, the $n$-th Bessel wave
  ($n=1,2, ...$)  is  stable when its frequency  is close enough to  $\Omega^{(n)}$, its  inception point.
 Our numerical analysis indicates that the Bessel wave loses its stability once its energy $E$ has grown above 
  the period-doubling bifurcation value. 
  A quadruplet of complex Floquet multipliers leaves the unit circle at this point signifying the onset of instability against an oscillatory mode with an additional frequency. 
  
While most of the clearly distinguishable, nonoverlapping,   spikes result from the 1:2 resonances with the Bessel waves,
some  correspond to the 1:3, 1:4 or 1:6 resonances. 
Similar to the 1:2 spikes, an exponentially localised solution at the base
of a 1:3, 1:4 or 1:6 projection has a core oscillating at the frequency $\omega= 2 \pi/T$
 and its second-harmonic  tail.
As this solution is continued up the slope of its spike, the contribution of higher harmonics to the  oscillation of the core and tail
 increases. Eventually the standing wave switches to the uniform regime where its core and tail oscillate at the same frequency --- $3\omega$, $4 \omega$ or $6\omega$.   The change of the temporal pattern is accompanied by the transformation of the spatial profile of the wave, 
 from the ``core-and-tail" composition  to a slowly decaying structure with no clearly defined core.
 
   It would be natural to expect this  weakly localised solution to merge with the 1/3, 1/4 or 1/6  
   undertone of a Bessel wave, implying the period multiplication of the latter.
   Numerically,  we do observe the bifurcations with $m=4$ and 6 while the period-tripling  of a Bessel wave is yet to be discovered.

\subsection{Higher resonances}

\begin{figure}[t]
 \begin{center} 
    \includegraphics*[width=\linewidth] {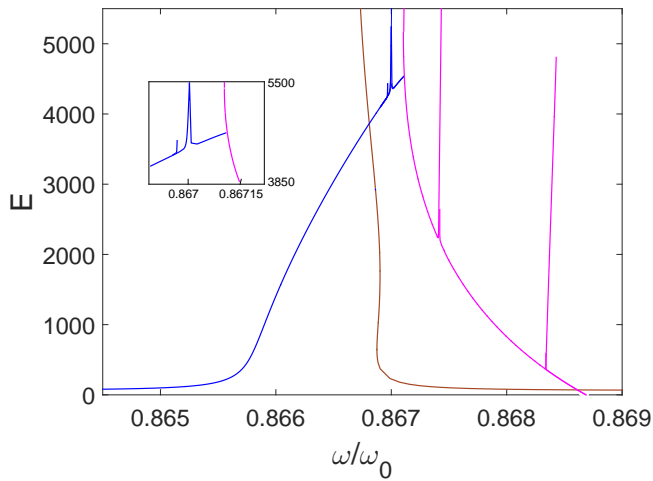}   
      \end{center}    
  \caption{\label{R100}  
The  bifurcation diagram in the neighbourhood of $\omega= \Omega^{(64)} /2$ in the ball of radius $R=100$.  
The inset zooms in on a tiny segment of the left slope of the primary peak (blue curve) that hosts  two baby spikes and 
 merges with the Bessel branch (shown in magenta). 
  }
 \end{figure}

 Fig \ref{R100} zooms in on the neighbourhood of  $\omega=  \Omega^{(64)}/2$ in the ball of radius $R=100$.
 Besides the primary spike pattern recognisable from our earlier  Fig \ref{doubling}, the diagram features several thinner vertical projections.
 These secondary, or  ``baby",  spikes result from   resonances with higher harmonics.

\begin{figure}[t]
 \begin{center} 
    \includegraphics*[width=\linewidth] {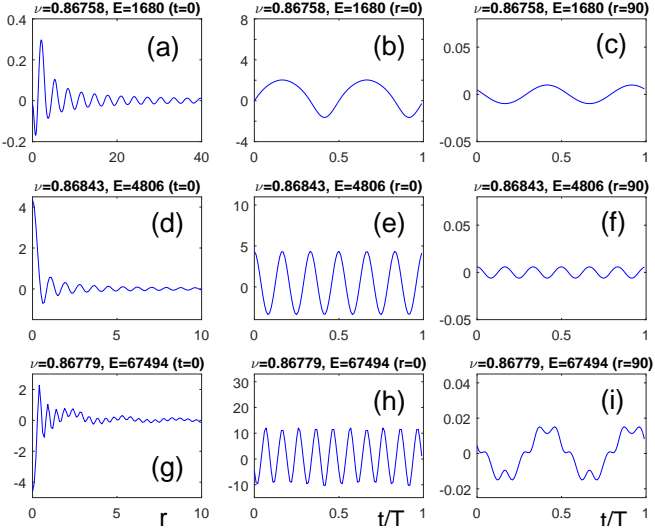}  
  \end{center}
  \caption{\label{2nd_Bessel} 
  Solutions on the Bessel branch and its two offshoots  in Fig \ref{R100}. 
  Top row:  the wave of frequency $2\omega$ found on the Bessel curve between the two baby spikes.
  Middle row: solution of frequency $6 \, \omega$ represented by the right-hand offshoot.
  Bottom row: solution of frequency $10 \, \omega$ corresponding to the left baby spike.
  }
 \end{figure}

 The magenta curve  in Fig \ref{R100} comprises the $1/2$ undertones of the Bessel wave.
 This branch and two needlelike secondary projections sprouting up from it   represent standing waves without clearly defined cores; see 
 Fig \ref{2nd_Bessel}. The top row in  Fig \ref{2nd_Bessel}
   corresponds to a solution occurring between the two  baby spikes; 
   it consists of  a pair of  identical cycles on the interval 
    $(0, 2\pi/\omega)$.
 The middle row  of  Fig \ref{2nd_Bessel}
 exemplifies   standing waves found on either slope of the ``lower"  baby spike
 (spike centred on $\omega/\omega_0=0.86834$).
 These include six repeated cycles. 
  As $E$ grows, both slopes of the ``lower" spike merge with the branch of the
  $1/6$ undertones of 
   {}{another  Bessel branch extending from $E=0$} (not shown in Fig \ref{R100}).
  Finally, 
 in the bottom row of Fig \ref{2nd_Bessel} we display a solution that belongs to the secondary projection appearing higher on the Bessel curve
 (spike centred on $\omega=0.86741$). 
 This coreless standing wave oscillates at the frequency  $10  \, \omega$.
 
 We note that solutions on both slopes of each of the two baby spikes emerging from the Bessel branch are stable.

\begin{figure}[t]
 \begin{center} 
    \includegraphics*[width=\linewidth] {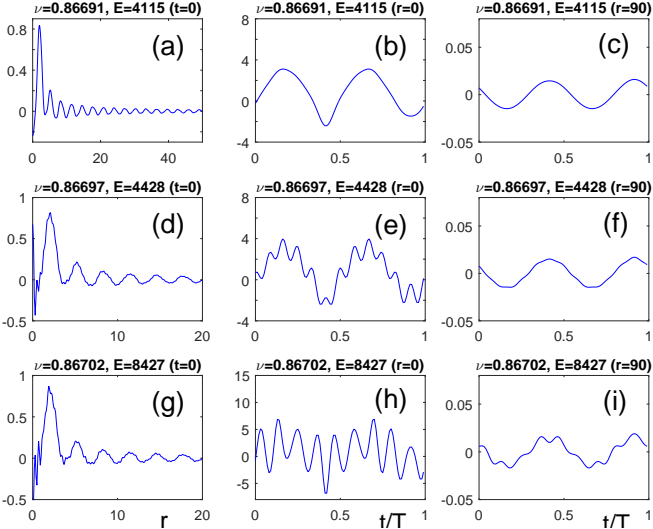}     
  \end{center}
  \caption{\label{left_curve}    Top row: solution of frequency $2\omega$
  on the blue side of the primary ``peak" shown in  
Fig \ref{R100}. 
Middle respectively bottom row: solutions of frequency $12 \,\omega$ respectively $9 \, \omega$ 
found on the left respectively right baby spike stemming out from the primary peak. 
(The two spikes are clearly visible in the inset to  Fig \ref{R100}).
 All three standing waves are coreless due to the proximity to the Bessel branch. }
 \end{figure}

 Fig \ref{left_curve} documents standing waves found on the left slope of the primary spike
 (the blue curve in Fig \ref{R100}) and secondary spikes emerging from it. The top row illustrates 
 the solution at a point of the primary curve near its merger with the Bessel branch.
 The structure  of this solution is similar to that in the bottom row of Fig \ref{6r}. 
 The wave does not have a clearly defined core while its central value $\left.\phi \right|_{r=0}$ 
 and a slowly decaying tail oscillate at the same frequency $2  \,\omega$.
 The middle and bottom rows in Fig \ref{left_curve} describe   solutions on the left and right baby spikes jutting out from the primary curve.
These have a large-amplitude $12 \, \omega$- and $9 \, \omega$-component, respectively.

 \subsection{Stability of standing waves}

With the  stability of the Bessel waves classified
earlier in this paper, we turn to the exponentially localised solutions comprising  the $E(\omega)$ curve in 
Fig  \ref{Ew}.

As we demonstrated in sections \ref{sta_small} and \ref{overpeak}, 
the monodromy matrix acquires a pair of real eigenvalues ($\zeta_1>1$ and $\zeta_2=  1/ \zeta_1$)  as the solution is continued 
over the peak at   $\omega_c = \omega_0 + \mu_c (2 \omega_0 R^2)^{-1}$ (the rightmost peak in Fig  \ref{Ew})
  in the direction of lower frequencies. The numerical analysis indicates that 
another real pair  ($\zeta_3>1$ and $\zeta_4=1/\zeta_3$)  leaves the unit circle  as $\omega$ is reduced past
 the  local energy minimum between the peak at $\omega_c$ and the next spike on its left. 
 
 Regardless of the choice of $R$,  real or complex  unstable  Floquet multipliers  persist over the entire interval $\omega_{\mathrm{min}} < \omega< \omega_c$,
where $\omega_{\mathrm{min}}$ is the point of
  minimum of the U-shaped envelope of the family of spikes. 
  For low energies, the instability is due to the real multipliers, $\zeta_1$ and $\zeta_3$. As  the solution ``climbs" up the energy slope,
  the real multipliers $\zeta_1$, $\zeta_3$,
  $1/\zeta_1$,  $1/\zeta_3$  merge, pairwise, and form a complex quadruplet.
  The quadruplet dissociates as the solution descends along the other slope of the same spike.

Stability properties in the region $\omega< \omega_{\mathrm{min}}$
prove to be  sensitive to the choice of $R$. The case of a small radius is exemplified by 
 the  ball of  $R=40$. Fig \ref{Ew}(b) depicts 
 the corresponding $E(\omega)$ diagram in
an interval of frequencies adjacent to 
 $\omega_0$. 
 (Note that the frequency  $\omega_{\mathrm{min}}$  is close  to the position of  the second spike from the right  in Fig \ref{Ew}(b).)

\begin{figure}[t]
 \begin{center}     
   \includegraphics*[width=\linewidth] {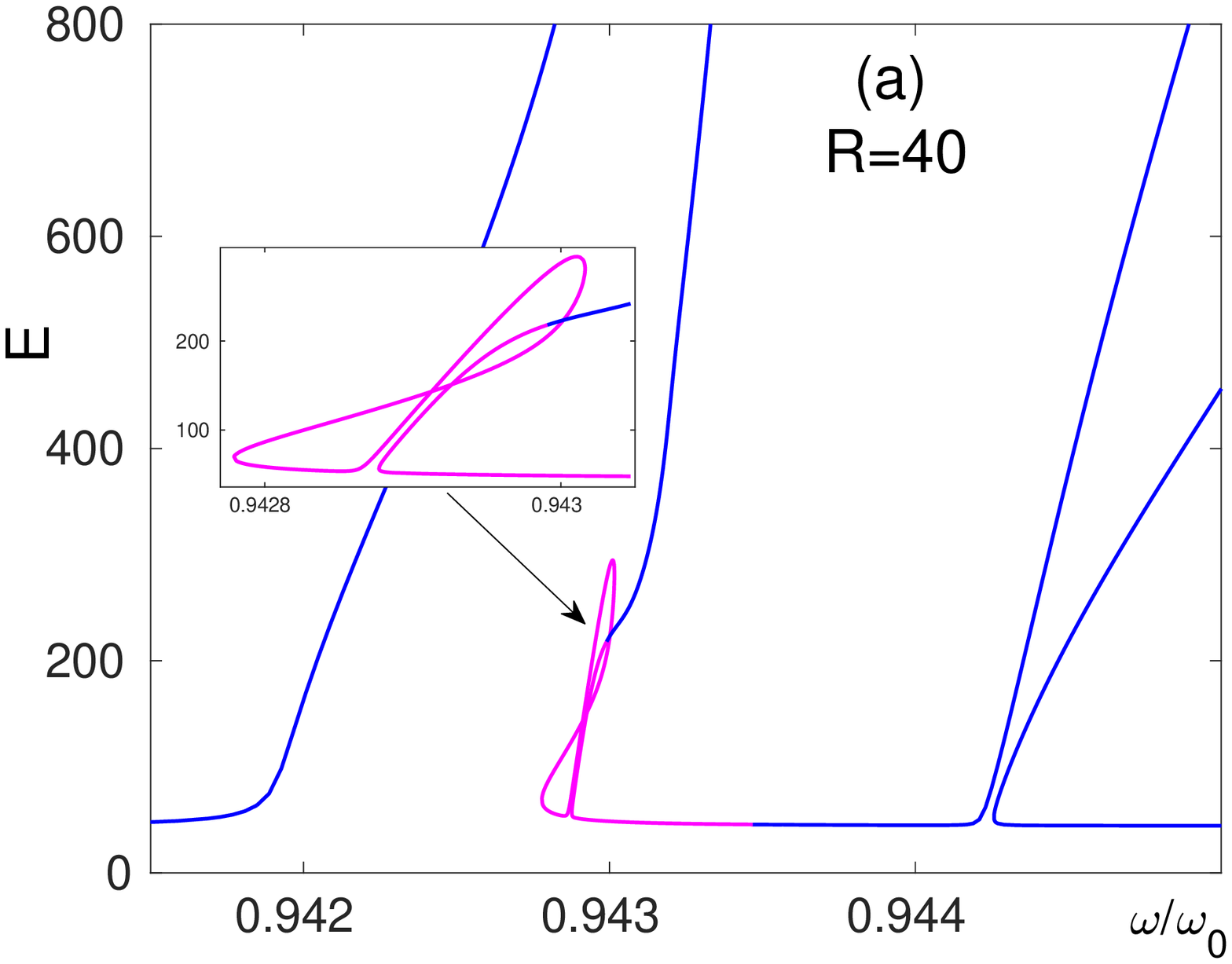}  
          \vspace*{6mm}
                     \includegraphics*[width=\linewidth] {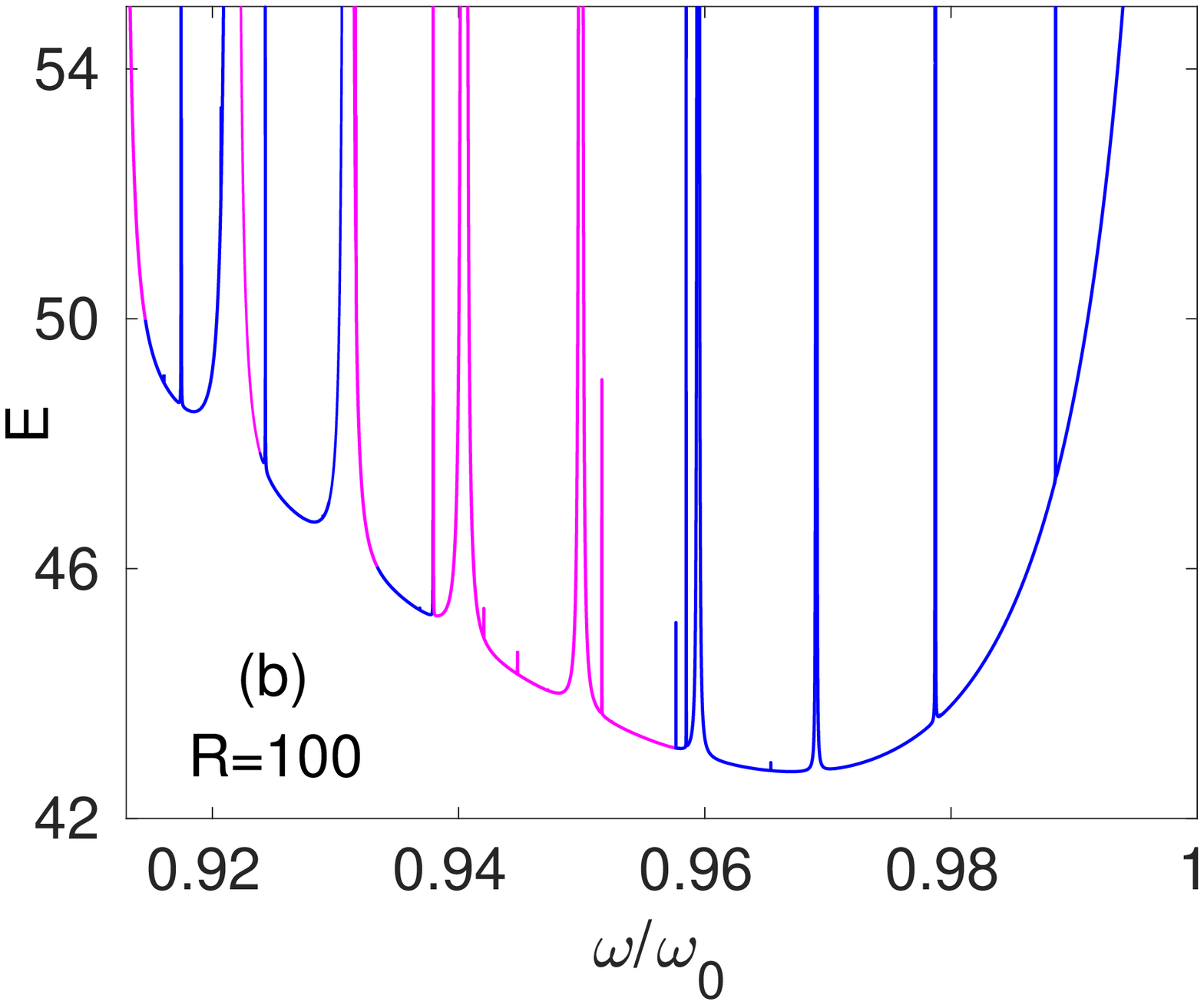} 
                                                                  \end{center}
  \caption{\label{stab40}  (a) The fine structure  of the
  third spike  from the right in
  Fig  \ref{Ew}(b). (Here $R=40$.) The spike is, in fact, a doublet;
  it  consists of two separate projections.  The inset zooms in on a
  figure-eight shaped isola occurring at the bottom  of the left ``subspike". 
  (b)  Stability of standing waves in  the ball of $R=100$, near the right end of its energy-frequency diagram (Fig \ref{Ew}(a)).   
    In (a) and (b) the magenta respectively blue  lines demarcate stable respectively unstable standing waves.
  }
 \end{figure}

 All frequencies between each pair of spikes  in  Fig \ref{Ew}(b)
 correspond to unstable solutions, with one or two pairs of real Floquet 
 multipliers off the unit circle. 
% (The multipliers are real,  $\zeta_1>1$ and $0<\zeta_2<1$.)
 %The left slope of the peak centred on $\omega \approx \omega_0$ as well as t
  The  second spike from the right (spike
 centred on   $\omega \approx 0.97$)
  is also entirely unstable. 
The only   intervals of stability   in Fig \ref{Ew}(b)
 are found at the base of the third and forth spike (centred on $\omega \approx 0.94$  and  $\omega \approx 0.92$, respectively).
Fig \ref{stab40}(a)   illustrates  stability of several branches associated with the third spike.

Turning  to a larger ball radius ($R=100$), the stability domain expands considerably.
 As  $\omega$  is reduced below
  $\omega_{\mathrm{min}}$ in that case, 
 two pairs of  real multipliers  form a complex quadruplet which, on further reduction,  converges 
  to two points on the unit circle.
  The value of $\omega$ at which the multipliers join the circle
  marks the beginning of a sizeable interval of
 stable frequencies 
  (Fig \ref{stab40}(b)). A continued reduction of $\omega$ sees an intermittent 
  appearance and disappearance  of one or several  complex quadruplets separating stability from instability intervals.

 \section{Concluding remarks} 
 \label{conclusions}

 {}{ 
 A linear standing wave in a ball results from the interference of an expanding spherical wavetrain of infinitesimal amplitude
 and the wavetrain reflected  from the ball's surface.  When continued  to finite amplitudes,  the resulting nonlinear solution does not have a well-defined core 
 and retains the $r$-dependence  similar to the spherical Bessel function $j_0(r)= \frac{\sin r}{r}$.  The total energy 
 associated with  this configuration  in a ball of radius $R$ is a multiple of  $R^2$.
 }

  {}{  
  A different type of nonlinear standing wave  in a ball is characterised by  an exponentially localised pulsating core. 
   The core is a fundamentally nonlinear feature; the nonlinearity shifts its frequency below the linear spectrum and this frequency shift 
 ensures the core's exponential localisation.   The core pulsating at the frequency $\omega$  radiates spherical waves with higher-harmonic frequencies $m \omega$, $m=2,3,...$.
  The standing pattern arises as a result of the interference of the expanding and reflected radiation wavetrains.}

  {}{ 
  As the radiation frequency $m \omega$
 comes near one of the  linear eigenfrequencies,  the solution approaches the corresponding  Bessel-like pattern. 
 The amplitude of the radiation increases and the total energy  in the ball of radius $R$ shoots up to values $O(R^2)$.
 By contrast, when $\omega$ is not near a resonant value,   the radiation from the core is weak.
  The standing            wave       in that case 
 may serve 
as an approximation to an {\it oscillon} --- a long-lived localised pulsating  structure in the infinite space ---
at the nearly-periodic stage of its evolution.
Nonlinear standing waves
provide information on the oscillon's energy-frequency relation and stability 
as well as topology  of the nearby regions of the phase space. 
 }

  {}{ 
  We examined the energy-frequency diagram of the standing wave 
and scrutinised 
the associated spatio-temporal transformation of the periodic  solution.  
 Results of this study can be summarised as follows.
 }

1. We have demonstrated the existence  of a countable set of 
standing waves (``Bessel waves") in a ball of a finite radius. 
The $n$-th  ($n=1,2, ...$) 
Bessel wave is a   solution of the boundary-value problem \eqref{2D}
with $n-1$ internal nodes in the interval $(0, R)$ 
and the envelope decaying in proportion to $r^{-1}$ as $r \to R$.
{}{The Bessel wave branches off the zero solution}
at $\omega=\Omega^{(n)}$; % where $\Omega^{(n)}$ is as in \eqref{G16}.
we have  constructed  it as an  expansion  in powers of the frequency detuning $\omega-\Omega^{(n)}$.
The Bessel wave remains stable in an interval of frequencies adjacent to $\Omega^{(n)}$.
%The perturbation expansion is; hence the 
% asymptotic solution is only valid in a small neighbourhood of each $\Omega^{(n)}$ ($n=1,2,....$).

2. The nodeless ($n=1$) Bessel wave is amenable to asymptotic analysis in a wider  frequency  range.
The pertinent asymptotic expansion is in powers of  $R^{-1}$ and the resulting solution 
is valid in a neighbourhood of $\omega_0$, the frequency of spatially-uniform oscillations.
This neighbourhood is found to be wide enough to include $\Omega^{(1)}$, the Bessel branch's inception point,
and $\omega_c$    ($\omega_c< \Omega^{(1)}$)  --- the frequency at which the energy curve $E(\omega)$ has a maximum. 
 The $n=1$ Bessel wave remains stable in the entire interval $\omega_c \leq \omega < \Omega^{(1)}$ but  loses its stability
 as $\omega$ is reduced below $\omega_c$.
 %These asymptotic conclusions have been  reproduced by the numerical analysis of the periodic boundary-value problem \eqref{2D}. 
% As the frequency is reduced further, 
% the $n=1$ Bessel wave transforms into a solution with an
%  exponentially localised core and a small-amplitude slowly decaying  second-harmonic  tail. 

 3.  The numerical continuation of the $n=1$  Bessel wave to values of $\omega$ below $\omega_c$  
% 3. The numerical continuation of the  solution with a localised core to lower values of $\omega$ 
produces an $E(\omega)$ curve with a sequence of  spikes
near the undertone points $\omega= \Omega^{(n)}/2$ with some large $n$. 
 The left and right slope of the spike adjacent to  $\frac{1}{2} \Omega^{(n)}$  result from a period-doubling bifurcation
of the $n$-th Bessel wave. In addition to the  primary sequence $\frac{1}{2} \Omega^{(n)}$, there are also thinner spikes near the $\frac{1}{3} \Omega^{(n)}$, $\frac{1}{4} \Omega^{(n)}$ and other undertones.
Slopes  of the spikes in the primary sequence 
host   secondary projections corresponding to higher resonances.

Away from the neighbourhoods of the spikes, the $E(\omega)$ curve follows a
 U-shaped arc with a single minimum at $\omega_{\mathrm{min}}=0.967 \omega_0$; the arc bounds all spikes from below.
 The arc is unaffected by the ball radius variations, as long as $R$ remains 
  large enough. This envelope curve
  describes the energy-frequency dependence of the nearly-periodic  oscillons in the infinite space.

 4.  Standing waves with energies lying on the envelope curve and at the base of the spikes 
 have an exponentially localised core and a small-amplitude slowly decaying  second-harmonic  tail. 
 We have classified stability of these solutions  against spherically-symmetric perturbations.
 Specifically,  we focused on the interval $0.91 \omega_0 < \omega < \Omega^{(1)}$ and considered two values of $R$:   $R=40$ and $R=100$.
 The ball of radius $R=40$ has only short stability intervals, located at the base of two spikes in its $E(\omega)$ diagram.
 By contrast, the   standing waves in the ball of $R=100$ have long stretches of stable frequencies.

Finally, it is appropriate to draw parallels  with resonance patterns observed in other systems.

The authors of Ref \cite{Morugante} carried out numerical continuations of breather solutions 
in a  one-dimensional necklace of Morse oscillators. Their $E(\omega)$ diagram features
resonances similar to those reported in section \ref{resonances} of the present paper.
Standing waves residing on the slopes of the spikes in our Figs \ref{Ew}, \ref{doubling}, \ref{R100} and \ref{stab40}(a)
are akin to the phonobreathers of Ref \cite{Morugante} while solutions represented by the U-arc in our Figs \ref{Ew}
correspond to their ``phantom breathers".  

A more recent Ref 
  \cite{Kev} is a numerical study of the circular-symmetric breathers in the 
 sine-Gordon 
 equation posed in a disc of  a finite radius. The  $E(\omega)$ diagram produced in that publication
 displays projections due to the odd-harmonic resonances.
 
 We note that neither Ref \cite{Morugante} nor \cite{Kev} observe a period-doubling  transmutation of phonon waves  into breathers.

\section*{Acknowledgments}
AB and EZ are grateful to the HybriLIT platform team for their
assistance with the {\it Govorun\/}  supercomputer computations. 
 This research was supported by the bilateral collaborative 
 grant from  the Joint Institute  for Nuclear Research and National Research Foundation of South Africa (grant 120467).


\section*{References}
\begin{thebibliography}{99}

\bibitem{Voronov1}
N A Voronov, I Y Kobzarev, and N B Konyukhova,
JETP Lett   {\bf 22}   290   (1975)


\bibitem{BM1}   I L   Bogolyubskii and   V G  Makhankov, JETP Lett {\bf 24} 12 (1976)


\bibitem{BM2} 
  I L      Bogolyubskii and      V G    Makhankov, JETP Lett {\bf 25} 107 (1977)

\bibitem{G1} M Gleiser, Phys Rev D {\bf   49} 2978 (1994)

\bibitem{CGM}
E J Copeland, M Gleiser and H-R M\"uller, Phys Rev D {\bf 52}  1920 (1995)

\bibitem{Riotto} A Riotto, Phys Lett B {\bf 365} 64 (1996)

\bibitem{Dymnikova} I. Dymnikova, L. Koziel, M. Khlopov, and S. Rubin,
Gravitation and Cosmology   {\bf 6}  311 (2000)



\bibitem{Broadhead} 
M. Broadhead and J. McDonald, Phys. Rev. D {\bf 72} 043519
(2005)


\bibitem{GInt} M Gleiser, Int. J. Mod. Phys. D {\bf 16} 219 (2007) 


\bibitem{11cosmo} E. Farhi, N. Graham, A. H. Guth, N. Iqbal, R. R. Rosales, and N. Stamatopoulos
Phys. Rev. D {\bf 77} 085019 (2008)



\bibitem{bubbling}
M. Gleiser, B. Rogers, and J. Thorarinson, Phys. Rev. D
{\bf 77} 023513 (2008)




\bibitem{Amin1} M. A. Amin,  arXiv:1006.3075 (2010)

\bibitem{Stamatopoulos}
M Gleiser, N Graham, and N Stamatopoulos,   Phys Rev D {\bf 83} 096010 (2011)



\bibitem{Amin2}
M.A. Amin, R. Easther, H. Finkel, R. Flauger and M.P. Hertzberg, Phys. Rev. Lett. {\bf 108}    241302  (2012) 


\bibitem{Zhou}
S-Y Zhou,    E J  Copeland,    R   Easther,    H  Finkel,   Z-G.Moua and P M  Saffin,
JHEP   {\bf 10}      026  (2013)




\bibitem{GG} M Gleiser and N Graham,
Phys Rev D {\bf  89} 083502 (2014)

\bibitem{Adshead}
P. Adshead, J. T. Giblin Jr., T. R. Scully and E. I. Sfakianakis, Journ of Cosmology and Astroparticle Physics,
{\bf 12} 034    (2015)


\bibitem{Bond} 
J R  Bond,     J Braden and L Mersini-Houghton,
Journ Cosmology and Astroparticle Physics    {\bf 09}   004   (2015)

\bibitem{Antusch}   S Antusch, F. Cefal\`a and S  Orani, Phys Rev Lett {\bf 118}   011303 (2017)






\bibitem{Hong}  
J-P Hong,    M Kawasaki,   and M  Yamazaki,
Phys Rev D {\bf 98} 043531 (2018)



\bibitem{LozAm}    K. D. Lozanov and M. A. Amin,   Phys. Rev. D { \bf99} 123504 (2019)

\bibitem{Cyn} 
D Cyncynates and  T Giurgica-Tiron,  Phys Rev  D {\bf 103} 116011 (2021)



\bibitem{GT2} M Gleiser and J Thorarinson, Phys Rev D {\bf 76} 041701(R)  (2007)

\bibitem{Achi}
V. Achilleos,  F. K. Diakonos, D. J. Frantzeskakis, G. C. Katsimiga, X. N. Maintas,  E. Manousakis, C. E. Tsagkarakis, and A. Tsapalis,
Phys Rev D {\bf 88}  045015   (2013)

\bibitem{Maslov} V. A. Koutvitsky and E. M. Maslov, Phys Rev D {\bf 83} 124028 (2011); Phys Rev D {\bf 102} 064007 (2020);
Phys Rev D {\bf 104} 124046 (2021)


% \bibitem{Zhang1}   H-Y Zhang, M A  Amin,  E J  Copeland,   P M Saffin and K D Lozanov, Journ  of Cosmology and  Astroparticle Physics    {\bf 07}    055    (2020)



\bibitem{Zhang2} H-Y Zhang, Journ  of Cosmology and
Astroparticle Physics    {\bf 03}    102    (2021)

\bibitem{Nazari}
Z Nazari,  M Cicoli,     K Clough
and F Muia,  Journ  of Cosmology and  Astroparticle Physics    {\bf 05}    027    (2021)


\bibitem{Kou1} 
X-X Kou, C Tian and
S-Y Zhou, Class. Quantum Grav. {\bf 38}   045005   (2021) 

\bibitem{Hira} 
T Hiramatsu,    E  I  Sfakianakis and M Yamaguchi, Journ High Energy Phys   {\bf  21} 2021 (2021)

\bibitem{Kou2} 
X-X Kou,  J B  Mertens,
 C Tian and
S-Y Zhou, Phys Rev D {\bf 105}   123505   (2022) 




\bibitem{Kolb}
E. W. Kolb and I. I. Tkachev, Phys. Rev. D {\bf 49} 5040 (1994)


\bibitem{Vaquero}  A Vaquero,  J Redondo and J Stadler,  Journ of Cosmology and Astroparticle Physics  {\bf 04} 012   (2019)


\bibitem{Kawa_axion}
M  Kawasaki,  W Nakanoa,
and E Sonomoto, Journ  of Cosmology and  Astroparticle Physics    {\bf 01}    047    (2020)


\bibitem{Olle} 
J Olle,  O Pujolas, and F Rompineve,   Journ  of Cosmology and  Astroparticle Physics    {\bf 02}    006  (2020)

\bibitem{Miyazaki}
M Kawasaki, K Miyazaki, K Murai,    H Nakatsuka,
 E  Sonomoto,  Journ  of Cosmology and  Astroparticle Physics {\bf 08}    066 (2022) 



\bibitem{string}
S Antusch, F Cefal\`a,  S Krippendorf, F Muia, S Orani and F Quevedo, JHEP {\bf 01} 083   (2018)


\bibitem{Sang} 
Y Sang and Q-G Huang,
Phys. Rev. D {\bf 100}  063516 (2019)

\bibitem{Kasu} S Kasuya, M Kawasaki, F Otani, and E Sonomoto,
Phys Rev  D   {\bf 102}  043016 (2020)


\bibitem{Farhi2}  E. Farhi, N. Graham, V. Khemani, R. Markov, R. Rosales, Phys. Rev. D 72 (2005) 101701(R);

 \bibitem{Graham} N. Graham, Phys. Rev. Lett. 98 (2007) 101801;   Phys. Rev. D 76 (2007) 085017;
 
 
  \bibitem{Gleiser4}   M Gleiser, N Graham, and N Stamatopoulos,    Phys Rev D {\bf 82}  043517 (2010);
%Sz. Borsanyi, M. Hindmarsh, arXiv:0809.4711 [hep-ph], 2008;



\bibitem{Sfakianakis} E. I. Sfakianakis, 	arXiv:1210.7568  (2012)



\bibitem{Hertz}
M.P. Hertzberg,  Phys. Rev. D {\bf 82} 045022    (2010)

\bibitem{Saffin} 
P. M. Saffin, P. Tognarelli, and A. Tranberg, J.  High Energy  Phys. {\bf 08} 125 (2014)




\bibitem{Borsanyi}  Sz. Bors\'anyi and M. Hindmarsh, Phys Rev D {\bf 79}  065010 (2009)


\bibitem{Kasuya}
S. Kasuya, M. Kawasaki, and F. Takahashi, Phys. Lett. B {\bf 559} 99 (2003)

\bibitem{Kawasaki} 
M Kawasaki,  F  Takahashi, and N Takeda, Phys. Rev.  D {\bf 92} 105024 (2015)

\bibitem{Mukaida} 
K Mukaida,  M Takimoto     and M Yamada,  JHEP {\bf 03} 122    (2017)

\bibitem{Ibe} 
M Ibe, M Kawasaki, W Nakano, and E Sonomoto,  Phys Rev  D {\bf 100} 125021 (2019);
 JHEP {\bf 04} 030  (2019)





\bibitem{Honda}  E P Honda and M  W Choptuik, Phys Rev D {\bf 65}  084037 (2002)

\bibitem{Gleiser10}
M
Gleiser  and M Krackow, Phys Lett B {\bf 80}5 135450  (2020) 


\bibitem{Fodor1} 
G Fodor, P Forg\'acz,  P Grandcl\'ement, and I R\'acz, Phys Rev D
{\bf 74}  124003 (2006)

\bibitem{GS1} M Gleiser and D Sicilia, Phys Rev Lett {\bf 101} 011602 (2008)

\bibitem{GS2}
 M Gleiser and D Sicilia, 
Phys Rev D {\bf 80} 125037 (2009)



\bibitem{G2}
M Gleiser, Phys Lett B {\bf 600} 126 (2004)

\bibitem{Fodor2}
G Fodor, P Forg\'acz,  Z Horv\'{a}th, \'{A} Luk\'{a}cs,  Phys Rev D
{\bf 78}  025003 (2008)


\bibitem{numerical_parameters}
For each $T$, equation  \eqref{A200} was substituted with  its    %  \eqref{ytt}
second-order finite-difference discretisation in $r$ and $t$. The numerical solutions 
in the ball of $R=40$ were obtained using the radial and temporal stepsizes $\Delta r= 0.08$ and 
$\Delta t= T/2000$, respectively. The continuation in the  $R=100$ and $R=150$ balls was carried out with
$\Delta r=0.1$ and $\Delta t=T/100$. The Floquet analyses used the forth-order Adams-Bashforth method for 
$R=40$ and the adaptive Runge-Kutta-Fehlberg algorithm for $R=100$ and $R=150$. 



\bibitem{Grimshaw}
{}{R Grimshaw. Nonlinear Ordinary Differential Equations. Applied Mathematics and Engineering Science Texts, 2. Blackwell Scientific Publications, Oxford, 1990}

\bibitem{Chicone} {}{C Chicone. Ordinary Differential Equations with Applications. Texts in Applied Mathematics, 34. Springer, New York, 2006}

\bibitem{Voronov2}
 N A Voronov  and I Y Kobzarev,   JETP Lett  {\bf 24}    532   (1976)

\bibitem{B}
I L Bogolyubskii, JETP Lett  {\bf 24} 579 (1976)



\bibitem{Kose} A M Kosevich and A S Kovalev,  Sov Phys JETP {\bf 40} 891 (1975)

\bibitem{Dashen} R Dashen, B Hasslacher, and A Neveu, Phys Rev D {\bf 11} 3424 (1975)





\bibitem{VK}
N.G. Vakhitov and  A.A. Kolokolov, Radiophys. Quantum Electron. {\bf 16}     783 (1973)
%M Grillakis, J Shatah and W Strauss, J. Funct. Anal. {\bf 74}   160   (1987);
\bibitem{Peli1} 
D.E. Pelinovsky, Proc. Roy. Soc. Lond. A  {\bf 461}   783 (2005)

\bibitem{Peli2} 
M. Chugunova and D. Pelinovsky, Journ Math Phys {\bf 51}  052901  (2010)

\bibitem{Anderson} {}{D L T Anderson and G H Derrick. Journ Math Phys {\bf 11}  1336 (1970) }


\bibitem{Morugante} 
A M Morgante, M Johansson, S Aubry and G Kopidakis, J  Phys  A: Math Gen {\bf 35}      4999 (2002) 


\bibitem{Kev}
P G Kevrekidis,   R  Carretero-Gonz\'alez,   J  Cuevas-Maraver,      D  J  Frantzeskakis,    J-G Caputo, B A  Malomed,
Commun Nonlinear Sci Numer Simulat {\bf 94}   105596 (2021) 

%  \bibitem{Fodor}   G Fodor, P Forg\'acs, P Grandcl\'ement, and I R\'acz,  Phys Rev D   {\bf 74} 124003 (2006)


%  \bibitem{Gleiser1}  M Gleiser, Phys Rev D {\bf 49} 2978 (1994)

%  \bibitem{Gleiser2}  E. J. Copeland, M. Gleiser, and H.-R. M\"uller  Phys. Rev. D  {\bf 52} 1920 (1995)


%  \bibitem{Farhi}   E. Farhi, N. Graham, V. Khemani, R. Markov, and R.  Rosales, Phys. Rev. D 72, 101701(R) (2005).





%  \bibitem{Honda} E. P. Honda and M.W. Choptuik, Phys. Rev. D 65, 084037   (2002).


\end{thebibliography}
\end{document}